\documentclass[11pt]{elsarticle}
\usepackage{bm}
\usepackage{graphicx}
\usepackage{amsmath}
\usepackage{cases}
\usepackage{amssymb}
\usepackage{booktabs}
\usepackage{multirow}
\usepackage{xcolor}
\usepackage{epstopdf}
\usepackage[normalem]{ulem}

\usepackage[figuresright]{rotating}

\usepackage[caption=false]{subfig}
\usepackage[top=1.1in, bottom=1.1in, left=1.0in, right=1.0in]{geometry}
\usepackage{tikz}
\usepackage{comment}
\usetikzlibrary{calc}

\begin{document}
	\title{A fast iterative scheme for the linearized Boltzmann equation}
	\author[strath]{Lei Wu~\corref{cor1}}
	\cortext[cor1]{Corresponding author.\\
		E-mail address: lei.wu.100@strath.ac.uk (L. Wu). }	
	
	\author[ed]{Jun Zhang}
	
	\author[xian]{Haihu Liu}
	
	\author[strath]{Yonghao Zhang }
	
	\author[ed]{Jason M. Reese}
	
	\address[strath]{James Weir Fluids Laboratory, Department of Mechanical and Aerospace Engineering, University of Strathclyde, Glasgow G1 1XJ, UK }
	
	\address[ed]{School of Engineering, University of Edinburgh, Edinburgh EH9 3FB, UK}
	
	\address[xian]{School of Energy and Power Engineering, Xi'an Jiaotong University, 28 West Xianning Road, Xi'an 710049, China}

	\begin{abstract}
		An iterative scheme can be used to find a steady-state solution to the Boltzmann equation, however, it is very slow to converge in the near-continuum flow regime. In this paper, a synthetic iterative scheme is developed to speed up the solution of the linearized Boltzmann equation, by penalizing the collision operator $L$ into the form $L=(L+N\delta{h})-N\delta{h}$, where $\delta$ is the gas rarefaction parameter, $h$ is the velocity distribution function, and $N$ is a tuning parameter controlling the convergence rate. The velocity distribution function is first solved by the conventional iterative scheme, then it is corrected such that the macroscopic flow velocity is governed by a diffusion equation which is asymptotic-preserving in the Navier-Stokes limit. The efficiency of the new scheme is verified by calculating the eigenvalue of the iteration, as well as solving for Poiseuille and thermal transpiration flows. We find that the fastest convergence of our synthetic scheme for the linearized Boltzmann equation is achieved when $N\delta$ is close to the average collision frequency. The synthetic iterative scheme is significantly faster than the conventional iterative scheme in both the transition and the near-continuum flow regimes. Moreover, due to the asymptotic-preserving properties, the SIS needs less spatial resolution in the near-continuum flow regimes, which makes it even faster than the conventional iterative scheme. Using this synthetic iterative scheme, and the fast spectral approximation of the linearized Boltzmann collision operator, Poiseuille and thermal transpiration flows between two parallel plates, through channels of circular/rectangular cross sections, and various porous media are calculated over the whole range of gas rarefaction. Finally, the flow of a Ne-Ar gas mixture is solved based on the linearized Boltzmann equation with the Lennard-Jones potential for the first time, and the difference between these results and those using hard-sphere intermolecular potential is discussed. 
	\end{abstract}	
	
	
	\maketitle

\section{Introduction}

The Boltzmann equation is fundamental to a broad range of  applications from aerodynamics to microfluidics~\cite{Cercignani2000}, and it is important to be able to solve it accurately and efficiently. Usually, the Boltzmann equation is solved by the stochastic Direct Simulation Monte Carlo (DSMC) technique, which uses a limited number of simulated particles to mimic the binary collisions and streaming of very large numbers of gas molecules~\cite{Bird1994}. In DSMC, the length of the spatial cell and the time step are required to be smaller than the local molecular mean free path and the mean collision time, respectively, and for this reason this technique becomes very slow and costly for near-continuum flows. Although time-relaxed and asymptotic-preserving Monte Carlo methods allow large time steps~\cite{Pareschi2001,JinShi2016}, the restriction on the size of the spatial cells has not yet been removed. The same problem, in fact, exists in deterministic numerical methods for the Boltzmann equation, where the streaming and collisions are treated separately in the splitting scheme~\cite{Tcheremissine2005,Filbet:2010la}.

The unified gas-kinetic scheme (UGKS) provides an alternative approach. It was first developed for the Bhatnagar-Gross-Krook (BGK) kinetic model~\cite{Bhatnagar1954,XuHuang2010}, then for the Shakhov model~\cite{Shakhov1968,Xu2012}, and finally generalized to the Boltzmann equation~\cite{LiuChang2016}. It handles the streaming and binary collisions simultaneously, so that for time-dependent problems, the time step is only limited by the Courant-Friedrichs-Lewy condition. Also, the UGKS retains the asymptotic-preserving property in the Navier-Stokes limit~\cite{chensongze}, so that the length of the spatial cells can be significantly larger than the molecular mean free path. Moreover, the UGKS is a finite volume method, and the analytical integral solution of the BGK-type model enables accurate flux evaluation at the cell interface, so that the essential flow physics can be captured with coarse grids~\cite{WangPeng2015}. These advanced properties make the UGKS a multiscale method for efficient and accurate calculations of rarefied gas flows over a wide range of the gas rarefaction. Recently, an implicit UGKS has been proposed which eliminates the time step limitation and further improves the numerical efficiency~\cite{zhuyajun2016}.

To find a steady-state solution to the Boltzmann equation, an iterative scheme is often adopted. In the free-molecular flow regime where binary collisions are negligible, the iterative scheme is efficient. This is because the gas molecules move in straight way so that any disturbance in one point can be quickly felt by all other points. However, for near-continuum flows the iterations slowly converge and the results are very likely to be biased by the accumulated rounding errors. Although the time and spatial steps can be large, the UGKS still needs a large number of iterations~\cite{zhuyajun2016}. This is governed by the underlying physics: the exchange of information through streaming becomes very inefficient when the binary collisions dominate. Therefore, it would be useful to develop an efficient numerical scheme to solve the Boltzmann equation that has the asymptotic-preserving property in the Navier-Stokes limit and converges to the steady-state rapidly.

Inspired by work on fast iterative methods for radiation transport processes~\cite{DSA2002}, accelerated iterative schemes have been developed for the linearized BGK and Shahkov models~\cite{Valougeorgis:2003zr} to overcome  slow convergence in the near-continuum flow regime. The fast iterative scheme is called a synthetic iterative scheme (SIS) since kinetic model equations are solved in parallel with diffusion equations for macroscopic quantities such as the flow velocity and heat flux. The SIS has been successfully applied to Poiseuille flow in channels with two-dimensional cross sections of arbitrary shapes~\cite{szalmas2010} using a BGK model for single-species gases, and flows of binary and ternary gas mixtures driven by the local pressure, temperature, and concentration gradients~\cite{szalmas2010,szalmas2016,szalmas2016_2} using the McCormack model~\cite{McCormack1973}. The fast convergence of the SIS is due to three factors: first, the macroscopic synthetic diffusion equations exchange the information very efficiently; second, the macroscopic flow quantities can be fed back into mesoscopic kinetic models; and third, macroscopic diffusion equations are solved more quickly than mesoscopic kinetic model equations.

In the present paper we aim to develop a SIS to solve the linearized Boltzmann equation (LBE) for Poiseuille and thermal transpiration flows. Although for the single-species LBE these canonical flows have been extensively studied for hard-sphere~\cite{Ohwada_sone_1989,Doi2010}, inverse power-law~\cite{lei_Jfm}, and even Lennard-Jones~\cite{Sharipov2009,Wu:2015yu} potentials, the numerical results for near-continuum flows are scarce. Moreover, for gas mixtures these flows have only been solved based on the hard-sphere model in a one-dimensional geometry~\cite{SiewertMixture2007}. We will calculate these flows through two-dimensional cross sections of arbitrary shape and investigate the influence of realistic intermolecular potentials for gas mixtures.

The rest of the paper is organized as follows. In Sec.~\ref{sectionII}, we briefly introduce the LBE for single-species gases and the conventional iterative scheme. Then by analyzing the SIS for the BGK model, we develop a SIS for the LBE and test its performance by calculating the eigenvalues of iterations and Poiseuille/thermal transpiration flows. We improve the efficiency of the proposed SIS by adjusting a parameter in the scheme, which can be determined prior to the numerical simulation. In Sec.~\ref{multiscale}, the SIS is applied to solve the rarefied gas flows in multiscale problems. In Sec.~\ref{tuneSec}, the SIS in the polar coordinates is proposed and numerical results of the LBE for Poiseuille flow through a tube are presented. In Sec.~\ref{sectionIII}, the SIS is extended to the LBE for gas mixtures, and Poiseuille flow of a Ne-Ar mixture is solved for the first time based on the Lennard-Jones potential. In Sec.~\ref{summary}, we conclude with a summary of the proposed numerical method and future perspectives.

\section{A synthetic scheme for the single-species LBE}\label{sectionII}

Consider the steady flow of a single-species monatomic gas through a channel of arbitrary cross section in the $x_1$-$x_2$ plane, subject to small pressure/temperature gradients in the $x_3$ direction. The velocity distribution function (VDF) can be expressed as $f=f_{eq}+h$, where
\begin{equation}
f_{eq}(\textbf{v})=\frac{\exp(-|\textbf{v}|^2)}{\pi^{3/2}}
\end{equation}
is the equilibrium VDF and $h(x_1,x_2,\textbf{v})$ is the deviated VDF satisfying $|h/f_{eq}|\ll1$. The LBE for $h$ reads: 
\begin{equation}\label{LBE}
v_1\frac{\partial
	{h}}{\partial{x_1}}+v_2\frac{\partial
	{h}}{\partial{x_2}}=L(h,f_{eq})+S,
\end{equation}
with the linearized Boltzmann collision operator~\cite{lei_Jfm}:
\begin{equation}\label{LBE_collision}
L=\underbrace{\iint B(|\textbf{v}-\textbf{v}_\ast|,\theta)  [f_{eq}(\textbf{v}')h({\textbf{v}}'_{\ast})+f_{eq}(\textbf{v}'_\ast)h({\textbf{v}}')-f_{eq}(\textbf{v})h({\textbf{v}}_\ast)]d\Omega d{\textbf{v}}_\ast}_{L^+}-\nu_{eq}(\textbf{v})h(\textbf{v}).
\end{equation}
In Eqs.~\eqref{LBE} and~\eqref{LBE_collision},  $\textbf{x}=(x_1, x_2, x_3)$ is the position vector normalized by the characteristic flow length $\ell$, $\textbf{v}=(v_1,v_2,v_3)$ is the molecular velocity vector normalized by the most probable speed $v_m=\sqrt{2k_BT_0/m}$ ($k_B$ is the Boltzmann constant, $T_0$ is the gas/wall temperature, and $m$ is the gas molecular mass),  $ B(|\textbf{v}-\textbf{v}_\ast|,\theta)$ is the collision kernel determined by the intermolecular potential~\cite{lei_Jfm, Wu:2015yu}, and 
\begin{equation}
\nu_{eq}(\textbf{v})=\iint B(|\textbf{v}-\textbf{v}_\ast|,\theta) f_{eq}(\textbf{v}_\ast)d\Omega{}d{\textbf{v}}_\ast
\end{equation}
is the equilibrium collision frequency.

. Finally, $S$ is the source term: 
\begin{equation}
S=\left\{
\begin{array}{r@{\;\;}l}
-X_Pv_3f_{eq}, & \text{\ \ \ for Poiseuille flow,}\\
-X_Tv_3(|\textbf{v}|^2-5/2)f_{eq}, & \text{\ \ \ for thermal transpiration flow,}
\end{array}
\right.
\end{equation} 
where $X_P$ and $X_T$ are the pressure and temperature gradients, respectively. For the LBE, since macroscopic quantities are proportional to $X_P$ and $X_T$,  we assume $X_P=X_T=-1$.

The macroscopic quantities of interest are the flow velocity normalized by the most probable speed:
\begin{equation}
U_3=\int{v_3h}d\textbf{v},
\end{equation}
the shear stresses normalized by equilibrium gas pressure $p_0$:
\begin{equation}
P_{13}=\int{2v_1v_3h}d\textbf{v}, \quad P_{23}=\int{2v_2v_3h}d\textbf{v},
\end{equation}
and the heat flux normalized by $p_0v_m$:
\begin{equation}
q_3=\int{\left(|\textbf{v}|^2-5/2\right)}v_3hd\textbf{v}.
\end{equation} 
The dimensionless mass flow rate $M$ and heat flow rate $Q$ are:
\begin{equation}
\begin{split}
\mathcal{M}=&\frac{1}{A}\iint U_3 dx_1dx_2, \\
\mathcal{Q}=&\frac{1}{A}\iint q_3 dx_1dx_2, 
\end{split}
\end{equation}
where $A$ is the area of the cross section.

The integro-differential system defined by Eqs.~\eqref{LBE} and~\eqref{LBE_collision} is usually solved by the conventional iterative scheme (CIS). Given the value of $h^{(k)}$ at the $k$-th iteration step, the VDF at the next iteration step is calculated by solving the following equation:
\begin{equation}\label{LBE_iteration}
\nu_{eq}h^{(k+1)}+v_1\frac{\partial
	{h}^{(k+1)}}{\partial{x_1}}+v_2\frac{\partial
	{h}^{(k+1)}}{\partial{x_2}}=L^+(h^{(k)},f_{eq})+S,
\end{equation}
where derivatives with respect to spatial variables are usually approximated by a second-order upwind finite difference. The process is repeated until relative differences between successive estimates of macroscopic quantities are less than a convergence criterion $\epsilon$. The number of iteration steps in CIS increases significantly when the ratio of the molecular mean free path to the characteristic flow length decreases, especially when the flow is in the near-continuum regime~\cite{Valougeorgis:2003zr}. It is our goal here to develop a fast iterative scheme to solve the LBE efficiently over the whole range of gas rarefaction.

\subsection{SIS for the BGK equation}\label{syn_BGK}

To begin with, we introduce the SIS for the BGK equation~\cite{Valougeorgis:2003zr}. The linearized Boltzmann collision operator in Eq.~\eqref{LBE_collision} is replaced by that of the BGK kinetic model, yielding the following equation for the deviated VDF $h$:  
\begin{equation}\label{BGK}
v_1\frac{\partial {h}}{\partial{x_1}} +v_2\frac{\partial {h}}{\partial{x_2}}=\underbrace{{\delta}[2U_3v_3f_{eq}-h]}_{L_{BGK}}+S,
\end{equation} 
where 
\begin{equation}\label{delta}
\delta=\frac{p_0\ell}{\mu{}v_m}
\end{equation}
is the rarefaction parameter, with $\mu$ being the gas shear viscosity.

Multiplying Eq.~\eqref{BGK} by the Hermite polynomials and applying the recursion relation, a set of  first-order partial differential equations can be obtained for various moments~\cite{Lihnaropoulos:2007gf}. Here, two relevant equations for the macroscopic flow velocity are listed:
\begin{eqnarray}
\frac{\partial U_3}{\partial x_1}=-\delta P_{13}-\frac{1}{4}\frac{\partial F_{2,0,1}}{\partial x_1}-\frac{1}{4}\frac{\partial F_{1,1,1}}{\partial x_2}, \label{Uz1} \\
\frac{\partial U_3}{\partial x_2}=-\delta P_{23}-\frac{1}{4}\frac{\partial F_{1,1,1}}{\partial x_1}-\frac{1}{4}\frac{\partial F_{0,2,1}}{\partial x_2}, \label{Uz2}
\end{eqnarray}
where
\begin{equation}
F_{m,n,l}(x_1,x_2)=\int h(x_1,x_2,\textbf{v})H_m(v_1)H_n(v_2)H_l(v_3) d\textbf{v}
\end{equation}
are the non-accelerated high-order moments, with $H_n(v)$ being the $n$-th order physicists' Hermite polynomial. The combination of Eqs.~\eqref{Uz1} and~\eqref{Uz2} leads to a diffusion equation for the flow velocity:
\begin{subequations}\label{diffusion_BGK}
	\begin{eqnarray}
	\frac{\partial^2 U_3}{\partial x_1^2}+\frac{\partial^2 U_3}{\partial x_2^2}&=&-\delta\left(\frac{\partial P_{13}}{\partial x_1}+\frac{\partial P_{23}}{\partial x_2}\right)
	-\frac{1}{4}\left(\frac{\partial^2 F_{2,0,1}}{\partial x_1^2}+2\frac{\partial^2 F_{1,1,1}}{\partial x_1\partial x_2}+\frac{\partial^2 F_{0,2,1}}{\partial x_2^2}\right) \label{no_shear_relation}\\
	&=&\left\{
	\begin{array}{r@{\;\;}l}
	-\delta
	-\frac{1}{4}\left(\frac{\partial^2 F_{2,0,1}}{\partial x_1^2}+2\frac{\partial^2 F_{1,1,1}}{\partial x_1\partial x_2}+\frac{\partial^2 F_{0,2,1}}{\partial x_2^2}\right) \label{shear_relation}, & \text{Poiseuille,}\\
	-\frac{1}{4}\left(\frac{\partial^2 F_{2,0,1}}{\partial x_1^2}+2\frac{\partial^2 F_{1,1,1}}{\partial x_1\partial x_2}+\frac{\partial^2 F_{0,2,1}}{\partial x_2^2}\right), & \text{thermal transpiration,}
	\end{array}
	\right.
	\end{eqnarray}
\end{subequations}
Note that in obtaining the final equation, we have used the relation ${\partial P_{13}}/{\partial x_1}+{\partial P_{23}}/{\partial x_2}=1$ for Poiseuille flow and  ${\partial P_{13}}/{\partial x_1}+{\partial P_{23}}/{\partial x_2}=0$ for thermal transpiration flow. The SIS for the BGK equation then works as follows~\cite{Valougeorgis:2003zr, Lihnaropoulos:2007gf}:
\begin{itemize}
	\item  When $h^{(k)}$ and $U_3^{(k)}$ are known at the $k$-th iteration step,  calculate the VDF $h^{(k+1)}$ by solving the following equation: \begin{equation}\label{BGK_iteration}
	{\delta}h^{(k+1)}+v_1\frac{\partial
		{h}^{(k+1)}}{\partial{x_1}}+v_2\frac{\partial
		{h}^{(k+1)}}{\partial{x_2}}=2{\delta}U_3^{(k)}v_3f_{eq}+S.
	\end{equation}
	
	\item From $h^{(k+1)}$, calculate the non-accelerated moments $F_{2,0,1}, F_{1,1,1}$, and $F_{0,2,1}$. 
	
	\item From $h^{(k+1)}$, calculate the flow velocity $U_3^{(k+1)}$ near the boundary.  However, for the flow velocity in the bulk (several computational layers away from the boundary), $U_3^{(k+1)}$ is obtained by solving the diffusion equation~\eqref{shear_relation}.

\end{itemize}

The above iterative procedure is continued until convergence. It should be emphasized that the relation ${\partial P_{13}}/{\partial x_1}+{\partial P_{23}}/{\partial x_2}=1 \ \text{or} \ 0$ for Poiseuille flow or thermal transpiration flow respectively is crucial for the fast convergence of the SIS. This is because non-accelerated moments are negligible at large values of the rarefaction parameter $\delta$, so the synthetic equation~\eqref{shear_relation} quickly adjusts the flow velocity to the solution of the Stokes equation, which is close to the solution of the linearized BGK equation. If the synthetic equation~\eqref{no_shear_relation} is used instead, with $P_{13}$ and $P_{23}$ calculated based upon the VDF obtained at each iteration step, the slow convergence at large values of $\delta$ is not improved, because it takes a lot of iterations to reach the condition ${\partial P_{13}}/{\partial x_1}+{\partial P_{23}}/{\partial x_2}=1\ \text{or} \ 0$; in the worst-case scenario, it may even lead to false convergence and wrong solutions when the spatial resolution is not high enough. However, Eq.~\eqref{shear_relation} guarantees the correctness of the solution at large values of $\delta$, as it has the asymptotic-preserving property in the Navier-Stokes limit~\cite{chensongze}.

\subsection{SIS for the LBE}

The development of a SIS for the LBE is not straightforward, since the linearized Boltzmann collision operator is much more complicated than that of the linearized BGK model. Directly following the method in Sec.~\ref{syn_BGK}, the following diffusion equation for the flow velocity is obtained:
\begin{equation}\label{LBE_false}
\begin{aligned}[b]
\frac{\partial^2 U_3}{\partial x_1^2}+\frac{\partial^2 U_3}{\partial x_2^2}=2\frac{\partial }{\partial x_1}\int v_1v_3Ld\textbf{v}+2\frac{\partial }{\partial x_2}\int v_2v_3Ld\textbf{v} 
-\frac{1}{4}\left(\frac{\partial^2 F_{2,0,1}}{\partial x_1^2}+2\frac{\partial^2 F_{1,1,1}}{\partial x_1\partial x_2}+\frac{\partial^2 F_{0,2,1}}{\partial x_2^2}\right),
\end{aligned}
\end{equation}
which, like Eq.~\eqref{no_shear_relation}, cannot improve the slow convergence at large values of $\delta$.  


To speed up the convergence, the relation ${\partial P_{13}}/{\partial x_1}+{\partial P_{23}}/{\partial x_2}=1 \ \text{or} \ 0$ for Poiseuille flow or thermal transpiration flow respectively must be reflected in the diffusion equation. For instance, as in Eq.~\eqref{shear_relation}, a term similar to $-\delta$ should appear on the right-hand-side of Eq.~\eqref{LBE_false} for Poiseuille flow. To achieve this, we penalize the Boltzmann collision operator by the BGK operator~\cite{Filbet:2010la}, i.e.,
\begin{equation}\label{BGK_penlty}
L=(L-L_{BGK})+L_{BGK},
\end{equation}  
and let 
\begin{equation}
\begin{aligned}[b]
2\int v_1v_3Qd\textbf{v}=2\int v_1v_3(L-L_{BGK})d\textbf{v}-\delta{P_{13}},\\
2\int v_2v_3Qd\textbf{v}=2\int v_2v_3(L-L_{BGK})d\textbf{v}-\delta{P_{23}}.
\end{aligned}
\end{equation} 
This transforms Eq.~\eqref{LBE_false} into
\begin{equation} \label{diffusion_LBE}
\begin{aligned}[b]
\frac{\partial^2 U_3}{\partial x_1^2}+\frac{\partial^2 U_3}{\partial x_2^2}=&-\delta
-\frac{1}{4}\left(\frac{\partial^2 F_{2,0,1}}{\partial x_1^2}+2\frac{\partial^2 F_{1,1,1}}{\partial x_1\partial x_2}+\frac{\partial^2 F_{0,2,1}}{\partial x_2^2}\right) \\
&+2\frac{\partial }{\partial x_1}\int v_1v_3(L-L_{BGK})d\textbf{v}+2\frac{\partial }{\partial x_2}\int v_2v_3(L-L_{BGK})d\textbf{v},
\end{aligned}
\end{equation}
which is very close to Eq.~\eqref{shear_relation} for the linearized BGK equation. At large values of the rarefaction parameter $\delta$, $\int(L-{}L_{BGK})v_1v_3d\textbf{v}$ and $\int(L-{}L_{BGK})v_2v_3d\textbf{v}$ approach zero, and Eq.~\eqref{diffusion_LBE} possesses the asymptotic-preserving property in the Navier-Stokes limit~\cite{chensongze}. Therefore, a SIS can be developed based on this equation. Note that for thermal transpiration flow, $\delta$ in Eq.~\eqref{diffusion_LBE} should be replaced by zero as ${\partial P_{13}}/{\partial x_1}+{\partial P_{23}}/{\partial x_2}=0$.

The SIS for the LBE then works as that for the BGK equations, with some slight changes:
\begin{itemize}
	\item When $h^{(k)}$ and $U_3^{(k)}$ are known at the $k$-th iteration step, we calculate $\int v_1v_3(L-L_{BGK})d\textbf{v}$ and $\int v_2v_3(L-L_{BGK})d\textbf{v}$. We also calculate the VDF $h^{(k+1/2)}$ by solving the following equation:
	\begin{equation}\label{syn_LBE}
	{\nu_{eq}}h^{(k+1/2)}+v_1\frac{\partial
		{h}^{(k+1/2)}}{\partial{x_1}}+v_2\frac{\partial
		{h}^{(k+1/2)}}{\partial{x_2}}=L^+(h^{(k)},f_{eq})+S.
	\end{equation}
	
	\item From $h^{(k+1/2)}$, we calculate the flow velocity $U^{(k+1/2)}$, and the non-accelerated moments $F_{2,0,1}, F_{1,1,1}$, and $F_{0,2,1}$.
	
	\item Near the boundary, we let $U_3^{(k+1)}=U_3^{(k+1/2)}$, while we solve the diffusion equation~\eqref{diffusion_LBE} to obtain the flow velocity in the bulk.
	
	\item A correction of the VDF is introduced in accordance with the changed flow velocity:
	\begin{equation}\label{guided}
	h^{(k+1)}=h^{(k+1/2)}+2(U_3^{(k+1)}-U_3^{(k+1/2)})v_3f_{eq}.
	\end{equation}
	
	\item The above steps are repeated until convergence.
\end{itemize}

Note that for the linearized BGK model~\cite{Valougeorgis:2003zr}, Eq.~\eqref{guided} is not necessary because the linearized collision operator at the next iterative step automatically changes when the flow velocity is corrected by the diffusion equation. In the LBE, however, the change of flow velocity does not directly change the collision operator at the next iterative step, so Eq.~\eqref{guided} is important.

\subsection{Numerical analysis of the convergence rate}\label{conv_analysis}

Analytical solutions for the eigenvalue $\omega$ are introduced in order to characterize the convergence rate of the iterative scheme for the linearized BGK equation~\cite{Valougeorgis:2003zr, Lihnaropoulos:2007gf}. However, this is more difficult for the LBE because of its intricate collision operator. Here we calculate the eigenvalue numerically in order to study the performance of both the SIS and the CIS. For simplicity, we consider a periodic system of length $\ell$ in the $x_1$ direction, while the system is homogeneous in the $x_2$ direction.

For the CIS~\eqref{LBE_iteration}, the VDF is solved in the following manner (hereafter in this section, $h$ and $U_3$ should be viewed as their Fourier transforms in the $x_1$ direction):
\begin{equation}\label{conv_iteration}
h^{(k+1)}=\frac{L^+(h^{(k)},f_{eq})+S}{\nu_{eq}+2i\pi{v_1}}, \ \ \ \ i=\sqrt{-1}.
\end{equation}
During iteration, the flow velocity $U_3^{(k+1)}=\int h^{(k+1)}v_3d\textbf{v}$ is recorded, and upon convergence the resultant series of the flow velocity are fitted by  $U_3(k)=U_{3\infty}+Ce^{-\lambda k}$. The eigenvalue $\omega$ is then calculated as $\omega=e^{-\lambda}$. It is obvious that the smaller $\omega$ is, the faster the convergence, and the case of $\omega=1$ means no convergence.

\begin{figure}[tbp]
	\centering
	\includegraphics[scale=0.58,viewport=0 0 580 380,clip=true]{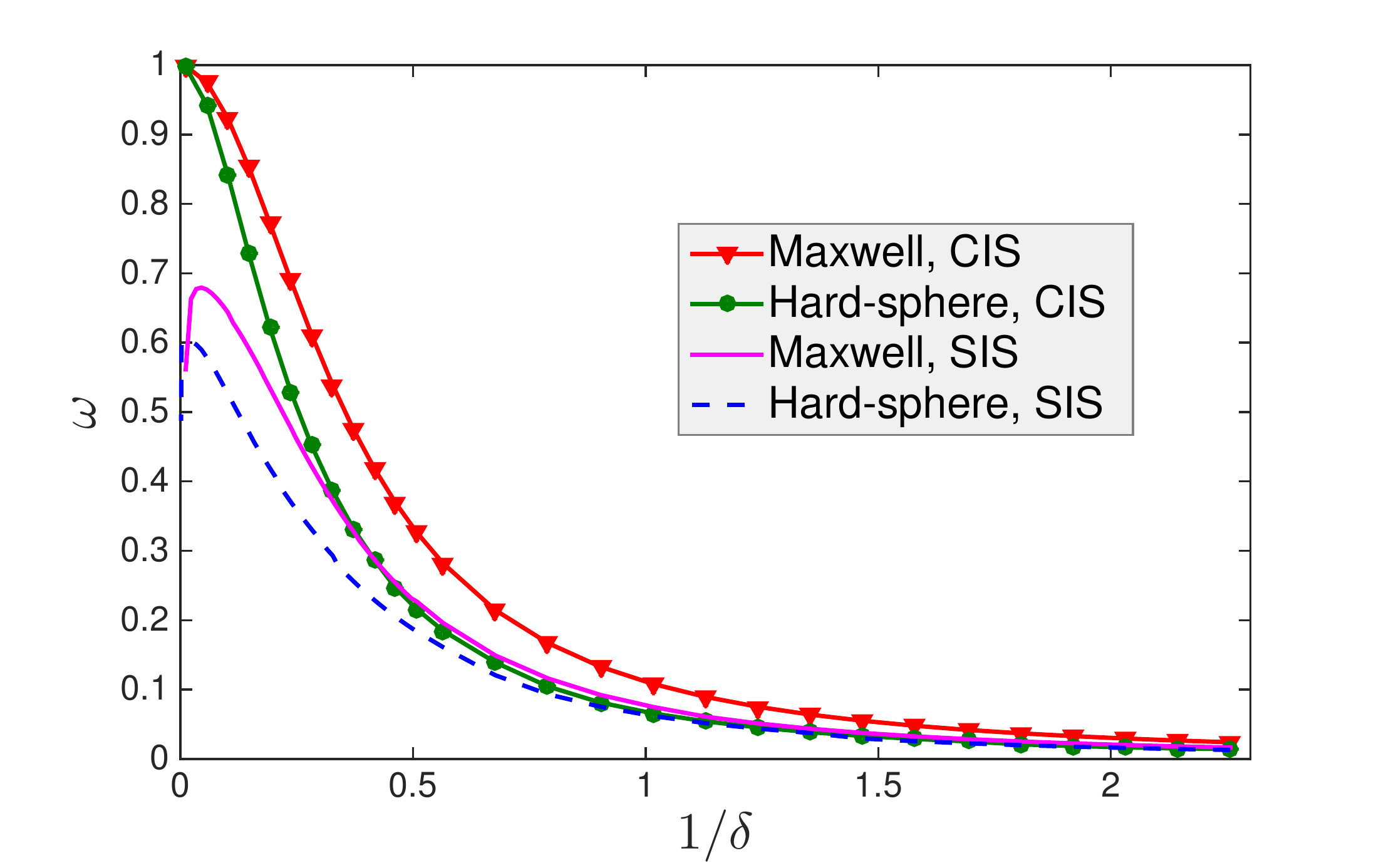}
	\caption{Eigenvalue $\omega$ versus the rarefaction parameter $\delta$ for the iterative scheme for the LBE with Maxwell and hard-sphere molecules (Note that Poiseuille and thermal transpiration flows have the same eigenvalue). Note that our method to calculate the eigenvalue for the SIS at large values of $\delta$ is not accurate, since the convergence is fast (it converges after one iteration) and we therefore have few data to find $\lambda$ through numerical fitting. However, the trend that $\omega$ for the SIS decreases with $1/\delta$ as $1/\delta\rightarrow0$ is clear.} 
	\label{eig_LBE}
\end{figure}

For the SIS, the VDF is first updated according to Eq.~\eqref{syn_LBE}: 
\begin{equation}\label{acceleratred_iteration}
h^{(k+1/2)}=\frac{L^+(h^{(k)},f_{eq})+S}{\nu_{eq}+2i\pi{v_1}}. 
\end{equation}
Then the flow velocity is calculated according to the diffusion equation~\eqref{diffusion_LBE} as
\begin{equation}
U_3^{(k+1)}=\left\{
\begin{array}{r@{\;\;}l}
(\delta-2i\pi{A_1}-4\pi^2A_2)/4\pi^2, & \text{\ \ \ Poiseuille,}\\
(-2i\pi{A_1}-4\pi^2A_2)/4\pi^2, & \text{\ \ \ thermal transpiration,}
\end{array}
\right.
\end{equation} 
where $A_1=2\int v_1v_3(L-L_{BGK})|_{h=h^{(k)}}d\textbf{v}$ and $A_2=\int h^{(k+1/2)}(2v_1^2-1)v_3d\textbf{v}$. Finally, this flow velocity is used to correct the VDF according to Eq.~\eqref{guided}: $h^{(k+1)}=h^{(k+1/2)}+2(U_3^{(k+1)}-U_3^{(k+1/2)})v_3f_{eq}$, where $U_3^{(k+1/2)}=\int{h^{(k+1/2)}}v_3d\textbf{v}$. The calculation of the eigenvalue for the SIS follows the same way as that for the CIS.

Figure~\ref{eig_LBE} presents the eigenvalues for both the SIS and the CIS. For small values of the rarefaction parameter $\delta$, both schemes have the same convergence rate. However, for large values of $\delta$, the CIS has extremely slow convergence $(\omega\approx1)$, while the SIS converges much faster. It is also interesting to note that the intermolecular potential greatly affects the convergence rate: at the same value of $\delta$, the solution of the LBE for hard-sphere molecules converges faster than that for Maxwell molecules\footnote{We assume the collision kernel $B(|\textbf{v}-\textbf{v}_\ast|,\theta)$ for Maxwell molecules is proportional to $1/\sqrt{\sin{\theta}}$, where $\theta$ is the deflection angle during binary collision, see Eq.~(2.3) in Ref.~\cite{lei_Jfm}.}, in both the SIS and the CIS.

\subsection{Numerical results for spatially-inhomogeneous systems}\label{NumSingle}

We now perform numerical simulations to demonstrate the efficiency and accuracy of the SIS for Poiseuille/thermal transpiration flows between infinite parallel plates and then through a two-dimensional square channel.

We first consider a gas flow between two infinite parallel plates located at $x_1=-1/2$ and $x_1=1/2$ (note that $x_1$ has been normalized by the distance between two parallel palates $\ell$).  Pressure and temperature gradients are applied in the $x_3$ direction only, so the flow is homogeneous in the $x_2$ direction and partial derivatives with respect to $x_2$ can be dropped. The discretization of the three-dimensional molecular velocity space, as well as the fast spectral approximation of the linearized Boltzmann collision operator, are given in Ref.~\cite{lei_Jfm}. We adopt the diffuse boundary condition for the gas-wall interaction. Due to symmetry, only half of the spatial region ($-1/2\le{}x_1\le0$) is simulated, with a specular-reflection boundary condition at $x_1=0$, and the diffuse boundary condition $h(v_2>0)=0$ at $x_1=-1/2$. The spatial domain is divided into 100 nonuniform sections, with most of the discrete points placed near the wall: $x_1=(10-15s+6s^2)s^3-0.5$, where $s=(0,1,\cdots,N_s)/2N_s$. The size of the smallest section is $1.24\times10^{-6}$, small enough to capture the Knudsen layer.

\begin{table}[tp]
	\centering
	\caption{Mass/heat flow rates in the Poiseuille flow of hard-sphere and Maxwell molecules between two parallel plates. Itr denotes the number of iteration steps to reach the convergence criterion $\epsilon=10^{-10}$. The results for the CIS are not shown at $\delta=100$ because it is hard to converge. }
	\begin{tabular}{p{0.5cm}p{0.30cm}p{0.80cm}p{0.80cm}p{0.80cm}p{0.80cm}p{0.80cm}p{0.20cm}p{0.80cm}p{0.80cm}p{0.80cm}p{0.80cm}p{0.80cm}cccccccccccc}
		\hline
		&  \multicolumn{6}{l}{Hard-sphere molecules} 
		&  \multicolumn{6}{l}{Maxwell molecules}   \\
		\cline{2-13}
		&  \multicolumn{3}{l}{SIS} 
		&  \multicolumn{3}{l}{CIS}  &  \multicolumn{3}{l}{SIS} 
		&  \multicolumn{3}{l}{CIS}   \\  \cline{2-13}
		$\delta$ & Itr & $\mathcal{M}$ & $-\mathcal{Q}$ & Itr & $\mathcal{M}$ & $-\mathcal{Q}$ 
		& Itr & $\mathcal{M}$ & $-\mathcal{Q}$ & Itr & $\mathcal{M}$ & $-\mathcal{Q}$ \\

		$0.01$ & 9  & 1.459 & 0.662 & 9     & 1.459 & 0.662  & 10 & 1.352 & 0.549  & 10 & 1.352 & 0.549 \\
		
		$0.05$ & 12 & 1.103 & 0.474 & 12    & 1.103 & 0.474  & 16 & 1.033 & 0.398  & 16 & 1.033 & 0.398  \\
	
		$0.1$  & 15 & 0.978 & 0.402 & 15    & 0.978 & 0.402  & 20 & 0.926 & 0.344  & 20 & 0.926 & 0.344  \\
		
		$0.5$  & 32 & 0.782 & 0.253 & 33    & 0.782 & 0.253  & 46 & 0.767 & 0.233  & 48 & 0.767 & 0.233  \\

		$1$    & 40 & 0.753 & 0.196 & 49    & 0.753 & 0.196  & 59 & 0.751 & 0.188  & 71 & 0.751 & 0.188  \\
		
		$2$    & 49 & 0.780 & 0.142 & 91    & 0.780 & 0.142  & 66 & 0.789 & 0.143  & 137 & 0.789 & 0.143  \\
	
		$5$    & 54 & 0.971 & 0.080 & 283   & 0.971 & 0.080  & 71 & 0.992 & 0.085  & 431 & 0.992 & 0.085  \\
		
		$10$   & 54 & 1.352 & 0.046 & 777   & 1.351 & 0.046  & 72 & 1.383 & 0.050  & 1183 & 1.382 & 0.050  \\
		
		$20$   & 55 & 2.154 & 0.024 & 2432  & 2.150 & 0.024  & 72 & 2.199 & 0.027  & 3684 & 2.194 & 0.027  \\
		
		$30$   & 54 & 2.968 & 0.017 & 4798  & 2.960 & 0.017  & 71 & 3.025 & 0.019  & 7245 & 3.017 & 0.019  \\				
	
		$50$   & 54 & 4.603 & 0.010 & 12038  & 4.582 & 0.010  & 72 & 4.686 & 0.012  & 18136 & 4.665 & 0.012  \\
		
		$100$  & 55 & 8.699 & 0.005 &   &  &                  & 73 & 8.848 & 0.006  & &  &   \\
		%
		%
		\hline
	\end{tabular}\par \label{table_poiseuille_1d_compare} 
\end{table}

For the one-dimensional problem, the shear stress is $P_{13}=x_1$ for Poiseuille flow and $P_{13}=0$ for thermal transpiration flow. The diffusion equation~\eqref{diffusion_LBE} is integrated to give the following first-order ordinary differential equation: 
\begin{eqnarray}\label{first_order}
\frac{\partial U_3}{\partial x_1}=-\delta P_{13}-\frac{1}{4}\frac{\partial F_{2,0,1}}{\partial x_1}+2\int v_1v_3(L-L_{BGK})d\textbf{v},
\end{eqnarray}
which is solved by a second-order upwind finite difference (with a first-order scheme at the wall), with the boundary condition $U_3(x_1=-1/2)=\int{v_3}h(x_1=-1/2)d\textbf{v}$ calculated from the VDF at each iteration. The iterations terminate when the relative errors in the mass and heat flow rates $(\mathcal{M}=2\int_{-1/2}^0{U_3}dx_1, \mathcal{Q}=2\int_{-1/2}^0{q_3}dx_1)$ between two consecutive iterations are less than $\epsilon=10^{-10}$.

\begin{table}
	\centering
	\caption{Mass/heat flow rates in Poiseuille/thermal transpiration flows of hard-sphere and Maxwell molecules along a channel of square cross section, as well as the number of iterations (Itr) to reach the convergence criterion $\epsilon=10^{-10}$ in the SIS.  }
	\begin{tabular}{clcclclcclc}
		\hline
		&  \multicolumn{5}{l}{Hard-sphere molecules} 
		&  \multicolumn{5}{l}{Maxwell molecules}   \\
		\cline{2-11}
		$\delta$ & Itr & $\mathcal{M}_P$ & $-\mathcal{Q}_P$ & Itr &  $\mathcal{Q}_T$ & Itr & $\mathcal{M}_P$  & $-\mathcal{Q}_P$ & Itr
		& $\mathcal{Q}_T$  \\

		$0.0$  & 3  & 0.419 & 0.210 & 3   & 0.944 & 3  & 0.419 & 0.210 & 3   & 0.944 \\
		
		$0.01$ & 5  & 0.413 & 0.205 & 5   & 0.924 & 6  & 0.411 & 0.201 & 5  & 0.918  \\

		$0.05$ & 6  & 0.402 & 0.194 & 6   & 0.882 & 7  & 0.398 & 0.188 & 8  & 0.869   \\
		
		$0.1$  & 7  & 0.395 & 0.186 & 7   & 0.847 & 9  & 0.391 & 0.179 & 9  & 0.832 \\
	
		$0.5$  & 13 & 0.379 & 0.153 & 13  & 0.695 & 18 & 0.378 & 0.149 & 21 & 0.677  \\
		
		$1$    & 16 & 0.382 & 0.132 & 18  & 0.589 & 19 & 0.382 & 0.130 & 29 & 0.572  \\
	
		$2$    & 24 & 0.400 & 0.106 & 21  & 0.458 & 33 & 0.405 & 0.108 & 40 & 0.443 \\
		
		$5$    & 30 & 0.484 & 0.068 & 29  & 0.275 & 39 & 0.494 & 0.072 & 48 & 0.265  \\
	
		$10$   & 31 & 0.644 & 0.042 & 31  & 0.162 & 40 & 0.659 & 0.045 & 50 & 0.157  \\
		
		$20$   & 32 & 0.981 & 0.023 & 32  & 0.088 & 40 & 1.002 & 0.026 & 53 & 0.086  \\
	
		$30$   & 31 & 1.323 & 0.016 & 27  & 0.061 & 40 & 1.349 & 0.018 & 52 & 0.059  \\				
		
		$50$   & 34 & 2.011 & 0.010 & 33  & 0.037 & 44 & 2.048 & 0.011 & 57 & 0.036  \\
	
		$100$  & 40 & 3.736 & 0.005 & 44  & 0.019 & 53 & 3.801 & 0.006 & 68 & 0.018  \\
		\hline
	\end{tabular}\par \label{table_poiseuille_2d_compare} 
\end{table}

A comparison between the SIS and the CIS is tabulated in Table~\ref{table_poiseuille_1d_compare} for Poiseuille flow. The relative differences in mass/heat flow rates between the two schemes is within 0.5\%, which demonstrates the accuracy of the SIS. The superiority of the SIS over the CIS is immediately seen: for the CIS, the number of iteration steps increase rapidly with the rarefaction parameter, while for the SIS it only slightly increases with $\delta$ in the free-molecular and transition flow regimes and saturates in the near-continuum flow regime $(\delta\ge10)$. Since, compared to the fast spectral approximation of the Boltzmann collision operator the time for solving Eq.~\eqref{first_order} is negligible, the CPU time saving is proportional to the time-step saving, and this is tremendous for the SIS. At $\delta=10$, the SIS is about 15 times faster than the CIS, while at $\delta=50$ it is about 220 times faster. 

It is interesting to note that for both the SIS and CIS, solutions of the LBE for hard-sphere molecules converge about 1.5 times faster than that for Maxwell molecules, a result which supports the convergence analysis in Sec.~\ref{conv_analysis}.

We also consider Poiseuille/thermal transpiration flows along a channel of square cross section. Due to symmetry, only one quarter of the spatial domain is simulated, which is divided into $50\times50$ non-uniform cells: in each direction, from the boundary to the center, the length of each cell side forms a geometric progression with a common ratio $1.05$. The diffusion equation~\eqref{diffusion_LBE} is discretized by a five-point central difference, and solved by the successive-over-relaxation method~\cite{LajosSzalmas2016}. Table~\ref{table_poiseuille_2d_compare} summarizes the numerical results from the SIS. The mass flow rate of thermal transpiration flow is not shown, as according to the Onsager-Casimir relation it is equal to the heat flow rate of Poiseuille flow. From this table it is seen that our SIS for the LBE works efficiently over the whole range of gas rarefaction. 



\begin{figure}[t]
	\centering
	{	\includegraphics[scale=0.5,viewport=10 0 460 360,clip=true]{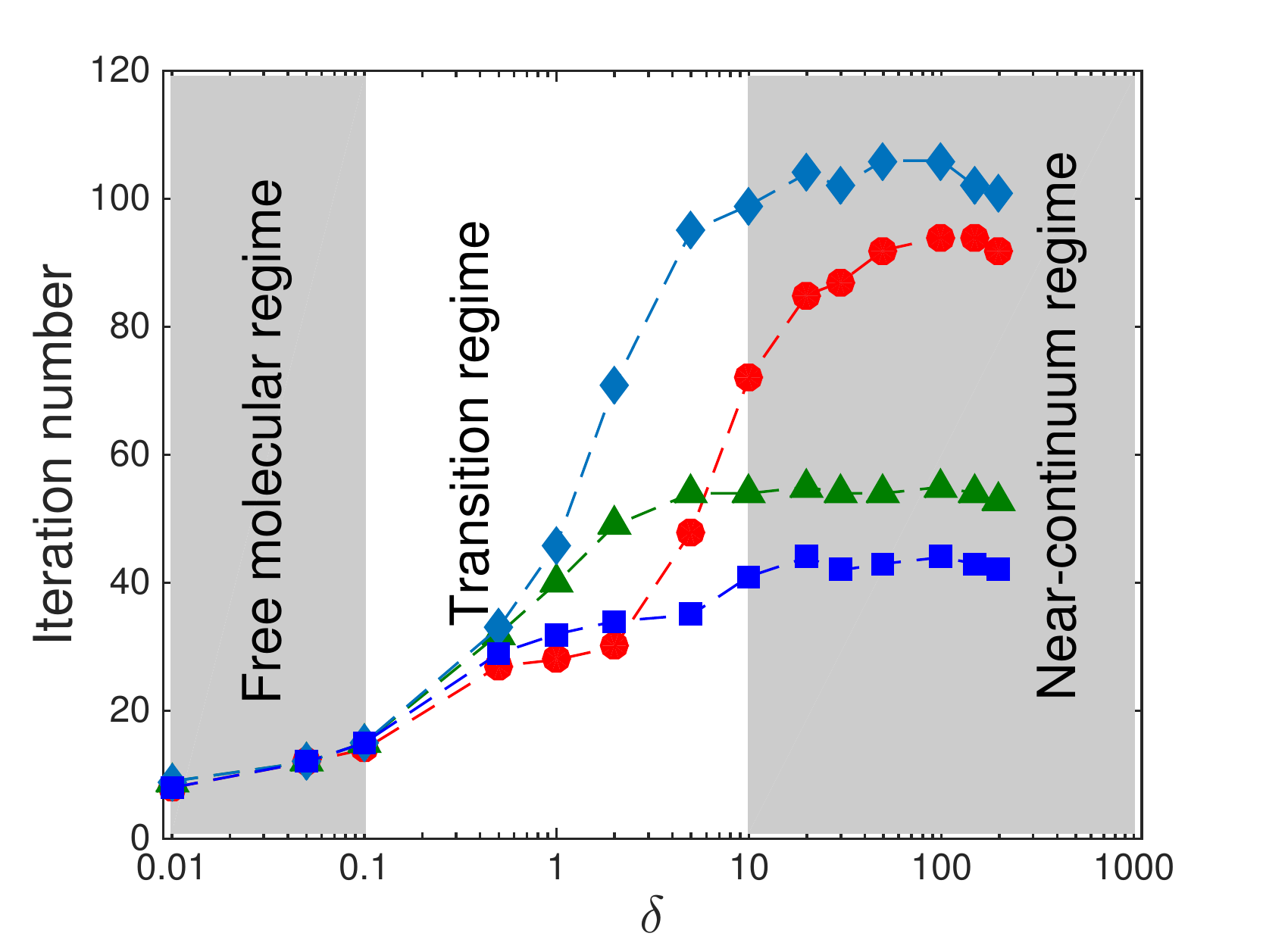}}
	{	\includegraphics[scale=0.5,viewport=10 0  460 360,clip=true]{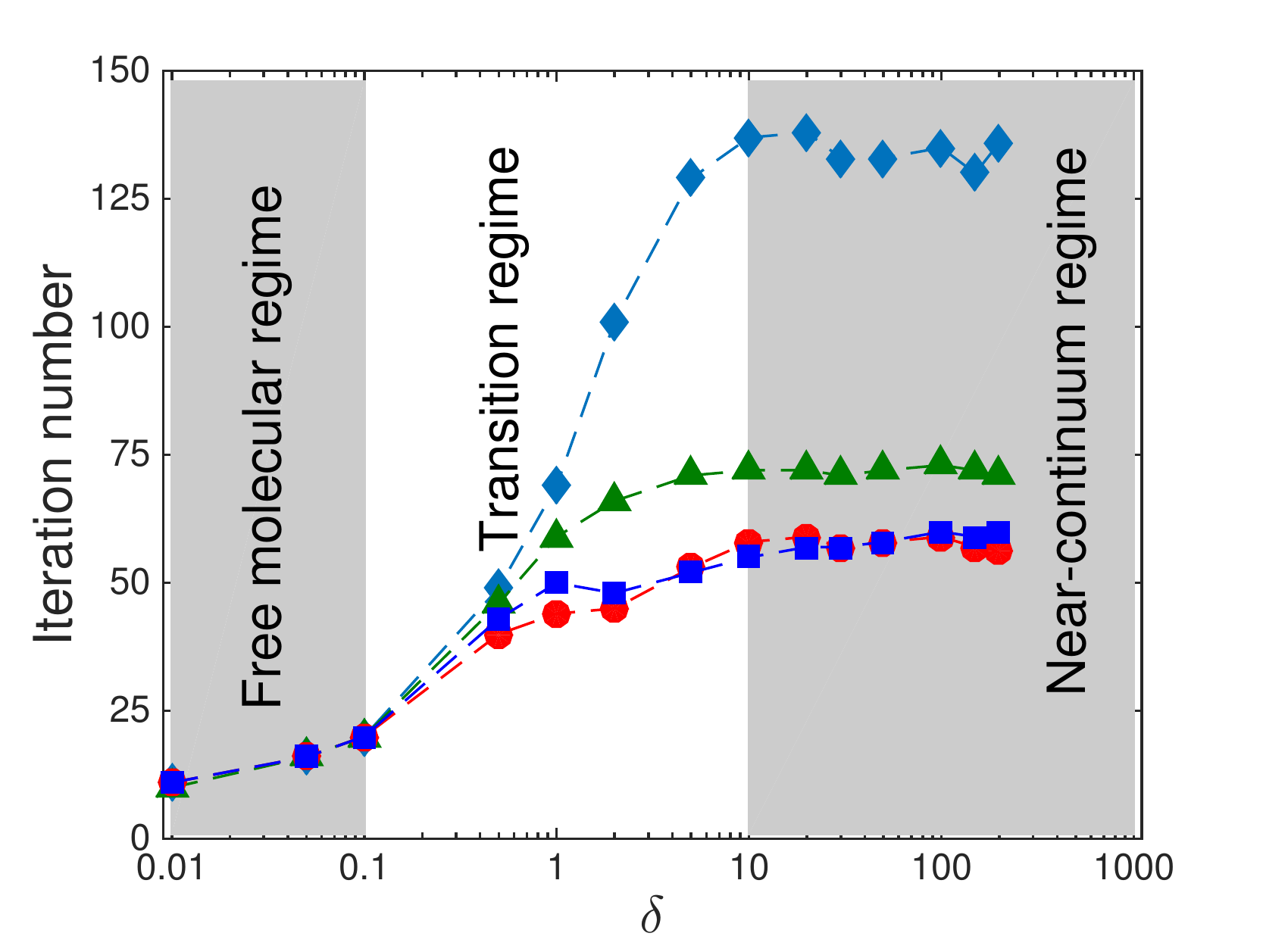}}
	\\
	\vskip 0.5cm
	{\includegraphics[scale=0.5,viewport=30 5 610 360,clip=true]{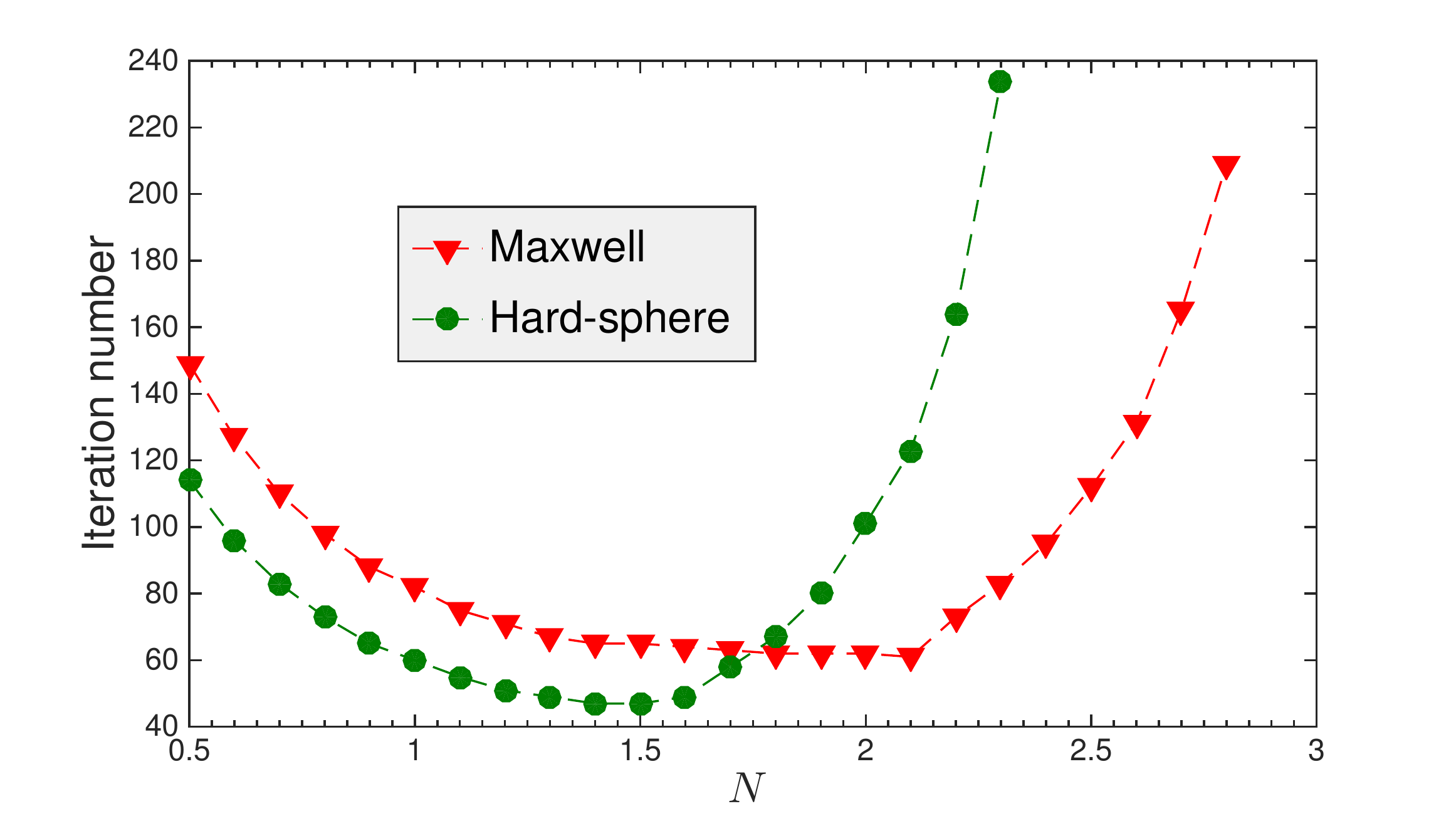}}
	\caption{Top row: iteration number versus the rarefaction parameter $\delta$ in the SIS (the convergence criterion is $\epsilon=10^{-10}$) for Poiseuille flow between two parallel plates. Left: hard-sphere molecules. Right: Maxwell molecules. Diamonds: $N=0.5$; Triangles: $N=1$; Squares: $N=1.5$; Circles: $N=2$. Bottom row: number of iterations to convergence versus $N$ in the SIS when the rarefaction parameter is $\delta=100$.   } 
	\label{Itr_num}
\end{figure}

\subsection{The most efficient scheme}

In Sec.~\ref{NumSingle} we saw that the SIS for the LBE can be faster than the CIS by several orders of magnitude in the near-continuum regime. Now we look at the possibility of speeding up the convergence even further, without modifying the scheme too much. To this end, we rewrite the linearized Boltzmann collision operator in the following form:
\begin{equation}\label{BGK_penlty2}
L=(L-NL_{BGK})+NL_{BGK},
\end{equation} 
where the constant $N$ is a tuning parameter which affects the convergence rate. 

With Eq.~\eqref{BGK_penlty} replaced by Eq.~\eqref{BGK_penlty2}, the diffusion equation~\eqref{diffusion_LBE} should be changed accordingly. Our numerical results for Poiseuille flow between two parallel plates show that, for a fixed $\delta$, all synthetic schemes with different values of $N$ converge to the same solution (with relative errors in flow rates less than $0.1\%$). However, the convergence rate varies with $N$. From the top row in Fig.~\ref{Itr_num} we see that in the free molecular regime all schemes have the same convergence rate, while in the transition regime the scheme with $N<1$ ($N>1$) converges slower (faster) than that with $N=1$. The situation becomes complicated in the near-continuum regime: for hard-sphere molecules, the case with $N=1.5$ converges fastest, followed by $N=1$, $2$, and $0.5$. For Maxwell molecules, however, the cases with $N=1.5$ and $N=2$ have roughly the same fast convergence, followed by $N=1$ and $0.5$. Similar behaviors are observed for the thermal transpiration flow.

To further investigate the relationship between the convergence iteration step and $N$ in the synthetic scheme, we fix $\delta=100$ and vary $N$. The numerical results in the bottom row of Fig.~\ref{Itr_num} show that the fastest convergence is achieved when $N$ is approximately 2 or 1.5 for Maxwell or hard-sphere molecules, respectively. This may be interpreted in terms of the average collision frequency. In the LBE, the equilibrium collision frequency $\nu_{eq}$ is in general a function of the molecular velocity, and the average collision frequency,
\begin{equation}
\bar{\nu}=\int\nu_{eq}(\textbf{v})f_{eq}(\textbf{v})d\textbf{v},
\end{equation}
varies between different intermolecular potentials even when the shear viscosity is the same. We found that for Maxwell and hard-sphere molecules, the average collision frequencies are 2.22 and 1.25 times larger than the rarefaction parameter, respectively, which are very close to the two values for the fastest convergence as shown in the bottom row of Fig.~\ref{Itr_num}. Therefore, to achieve the best performance of the synthetic scheme we suggest using
\begin{equation}
N=\frac{\bar{\nu}}{\delta}.
\end{equation}

\subsection{Further benefit of using SIS}

In addition to the significant speed up of convergence, the SIS can also help to reduce the spatial resolution. It is well-known that to solve the kinetic equations, the size of the spatial cells in the traditional discrete velocity method should be smaller than the molecular mean free path\footnote{Remarkably, the UGKS does not have this restriction.}, so that numerical results are reliable when the artificial viscosity is much smaller than the physical viscosity. In the SIS, the macroscopic flow velocity is obtained by solving the diffuse equation~\eqref{diffusion_LBE}, so the spatial resolution can be relatively coarser.


\begin{figure}[t]
	\centering
	{	\includegraphics[scale=0.65,viewport=30 0 620 380,clip=true]{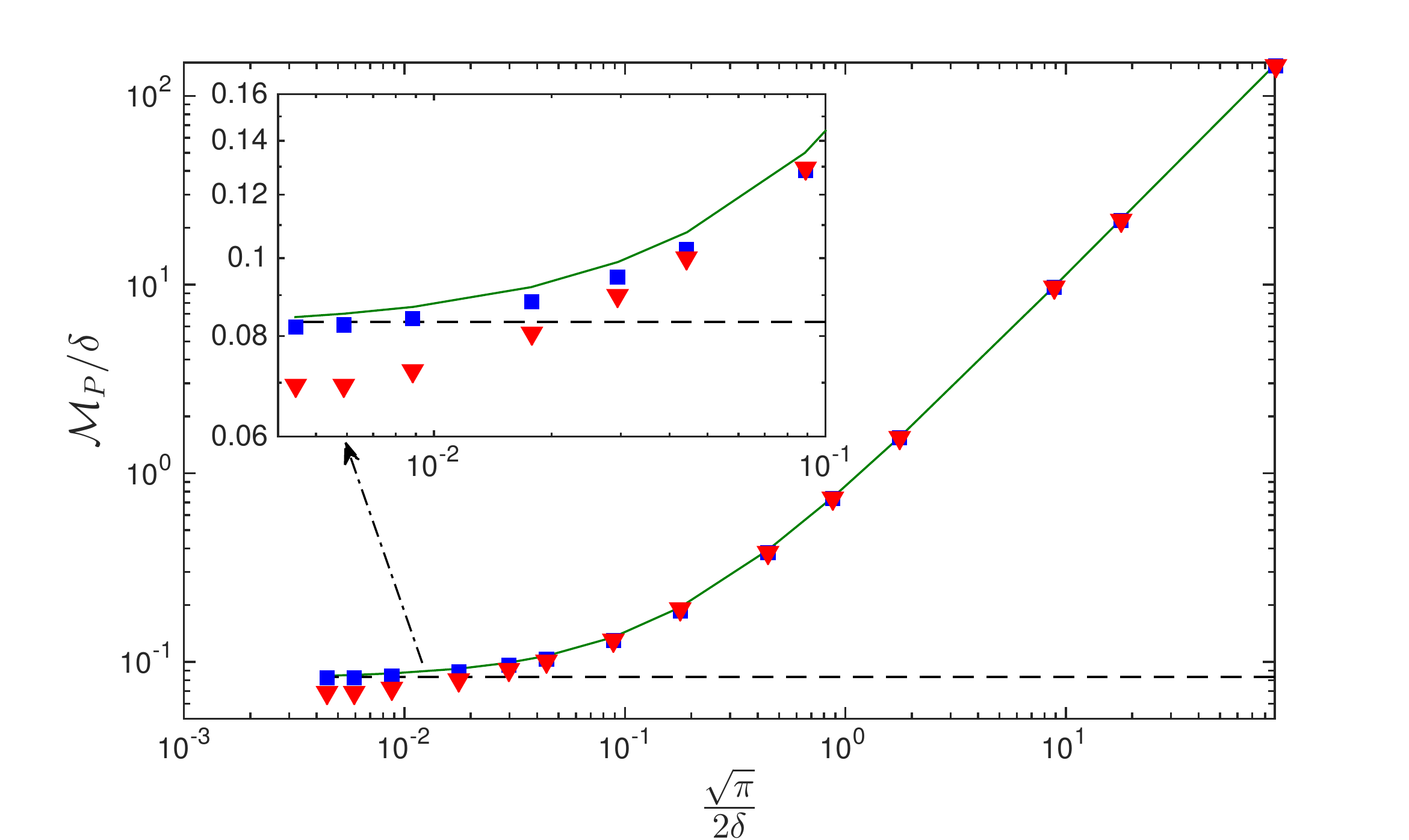}}
	\caption{The apparent gas permeability in the Poiseuille flow of the hard-sphere gas between two parallel plates. Solid lines: ``reference'' solution obtained from Table~\ref{table_poiseuille_1d_compare}. Squares and triangles: the accelerated and non-accelerated solutions of the LBE when the half spatial region is divided into 10 uniform sections, respectively. Dashed lines: the intrinsic permeability $\kappa=1/12$ when $1/\delta=\infty$.   } 
	\label{permeability}
\end{figure}

To demonstrate this, we run the test case in Sec.~\ref{NumSingle} again, but the half spatial domain is now divided into 10 uniform cells. Both the accelerated and non-accelerated schemes on this coarse grid are compared to the ``reference'' solutions in Table~\ref{table_poiseuille_1d_compare}. Since the mass flow rate in Poiseuille flow increases rapidly with $\delta$, we study how the apparent gas permeability  change with the rarefaction parameter. Here, the apparent gas permeability, which is normalized by $\ell^2$, is defined as
\begin{equation}
\kappa=\frac{\mathcal{M}_P}{\delta}.
\end{equation}
Note that, according to the Navier-Stokes equation with the no-slip velocity boundary condition, $\kappa=1/12$ when $\delta\rightarrow\infty$; this permeability is also known as the intrinsic or liquid permeability. The apparent gas permeability is  always larger than the intrinsic permeability, and increases with $1/\delta$ (or the Knudsen number).

Figure~\ref{permeability} shows the apparent gas permeability obtained at different spatial resolution and rarefaction parameter. It is clear that the SIS with the coarse spatial resolution can yield good results, while the non-accelerated iterative scheme has larger errors at large values of $\delta$. For instance, when $\delta$=150, the non-accelerated scheme underpredicts the apparent gas permeability by about 12.5\%. This error continues to increases with $\delta$: we tested the BGK model when $\delta=10^4$ and found that the relative error is about 62.5\%.

\section{SIS in multiscale problems}\label{multiscale}

In this section, we investigate the performance of the SIS in more complex geometries, where the problems are multiscale in the sense that the rarefaction parameter varies by several orders of magnitude if different characteristic size of solids is chosen, especially when the geometry is a fractal.

\subsection{Rarefied gas flow through the Sierpinski carpet}

\begin{figure}[t]
	\centering
	\subfloat[Level 1]	{	\includegraphics[scale=0.4,viewport=50 0 360 310,clip=true]{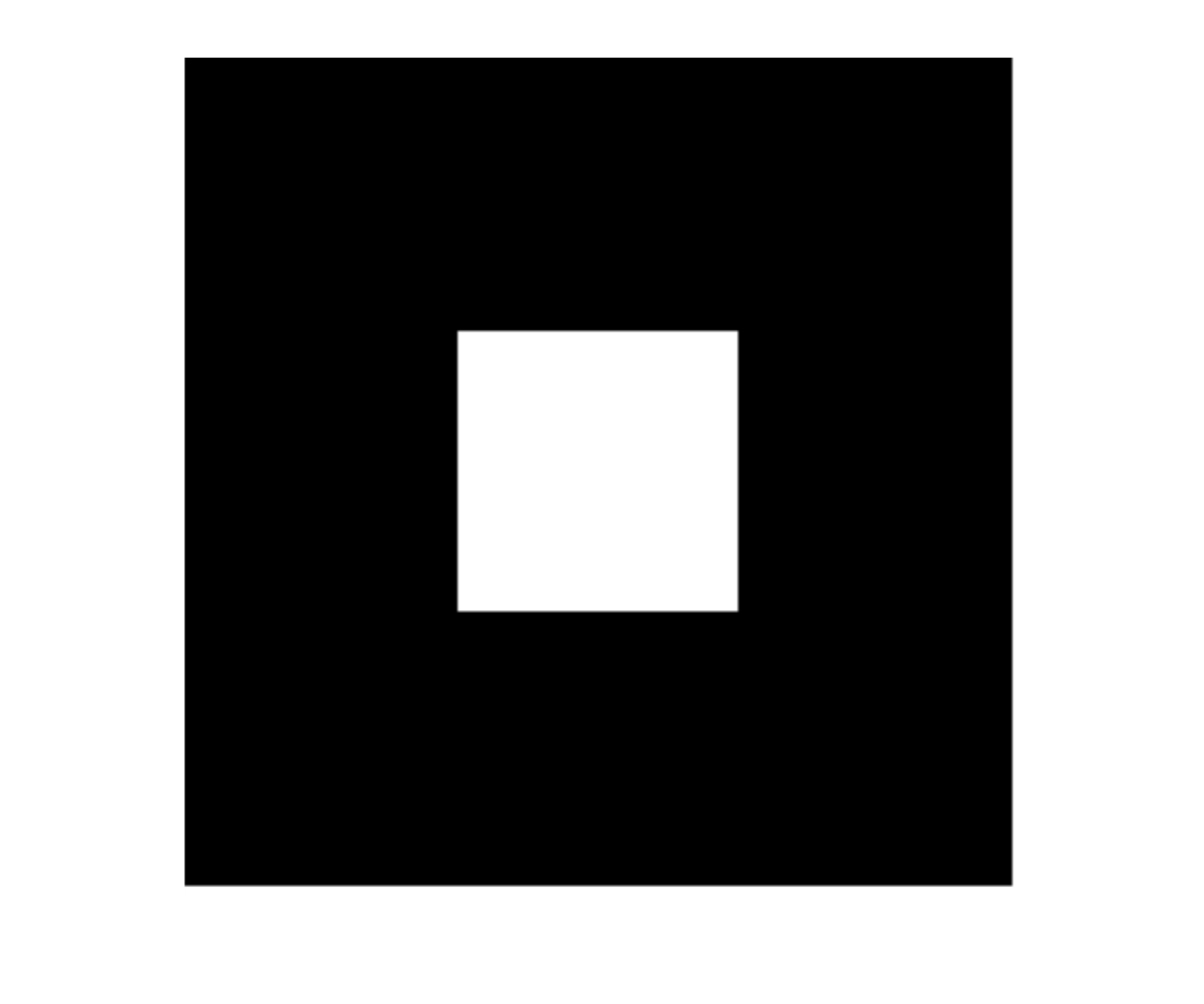}}
	\hskip .5cm
	\subfloat[Level 2]	{	\includegraphics[scale=0.4,viewport=50 0 360 310,clip=true]{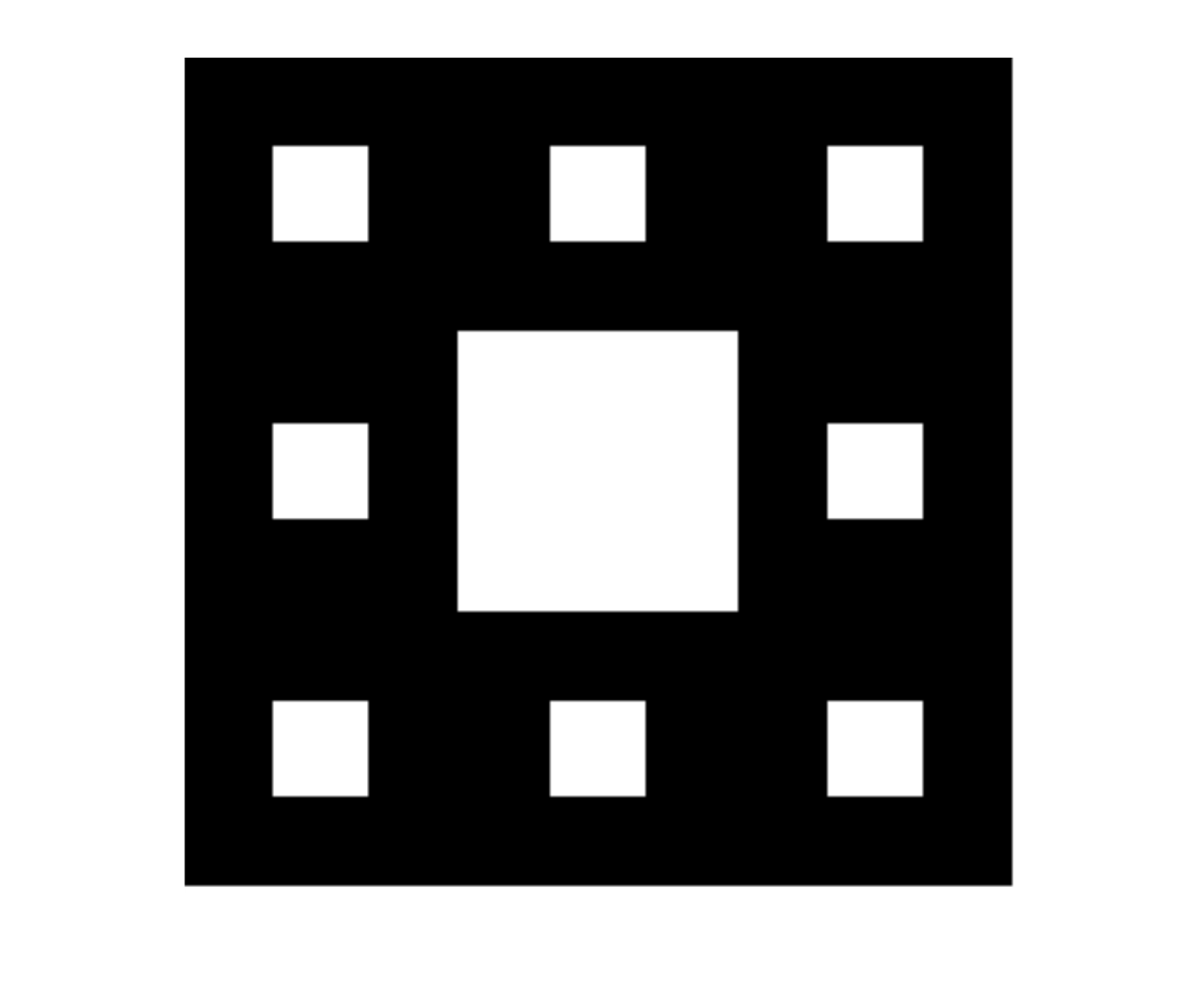}}
	\hskip .5cm
	\subfloat[Level 3]	{	\includegraphics[scale=0.4,viewport=50 0 360 310,clip=true]{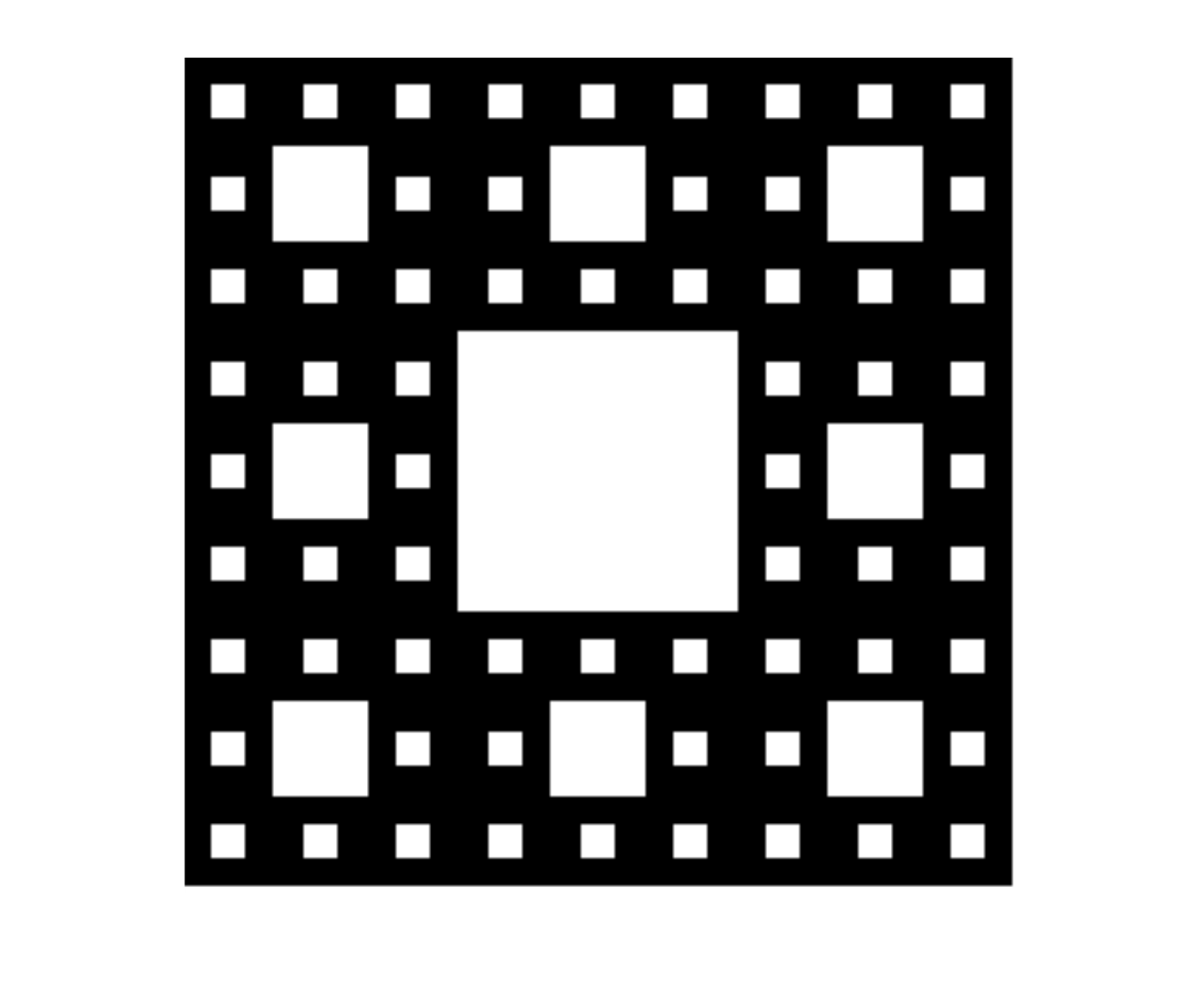}}
	\caption{The Sierpinski carpet generated at different levels of recursion. White regions represent the solid, while the gas can flow through the black regions.  } 
	\label{Sierpinski}
\end{figure}

\begin{figure}[thb]
	\centering
	{	\includegraphics[scale=0.7,viewport=150 60 830 370,clip=true]{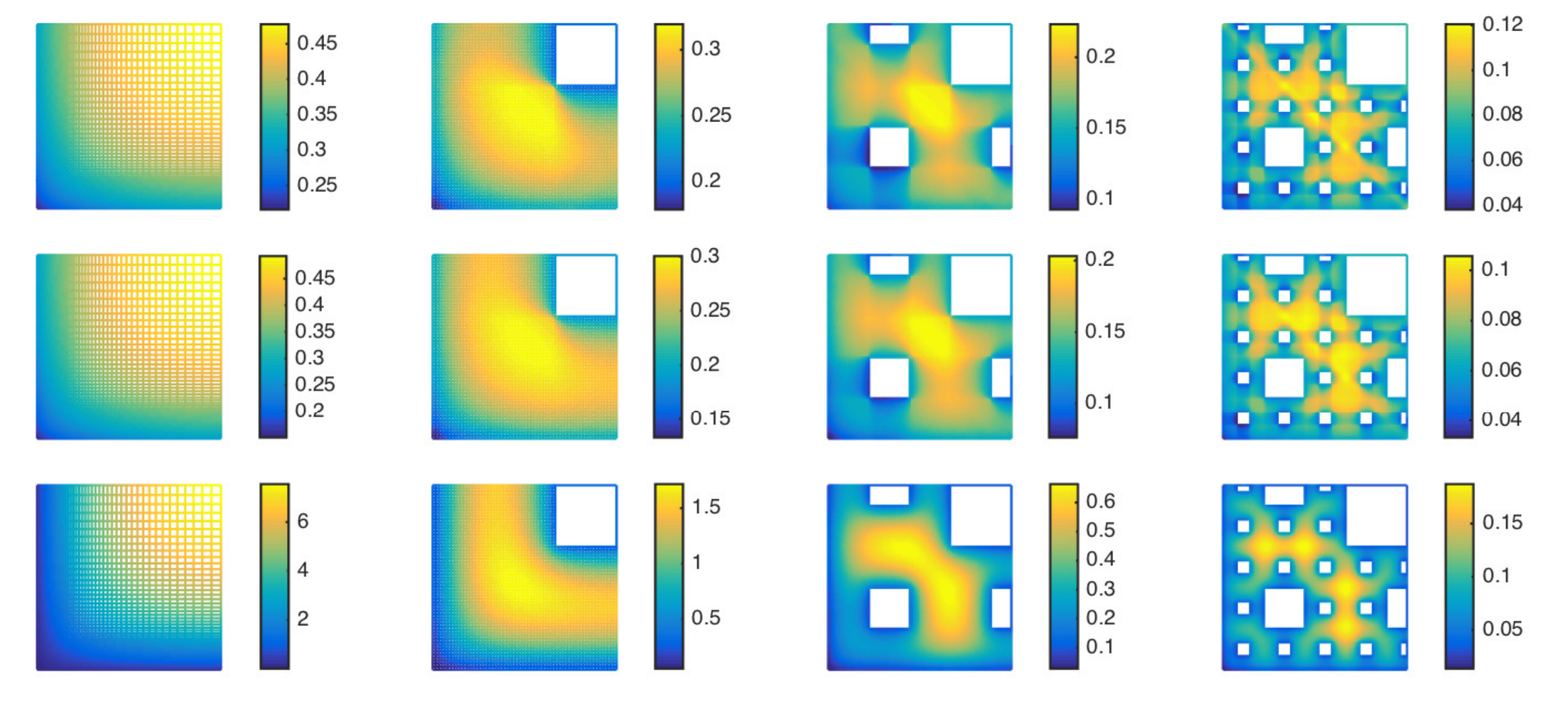}}
	\caption{Velocity contours in the Poiseuille flow of the hard-sphere gas through the Sierpinski carpets generated at different levels of recursion, when $\delta=1$ (top row), 10 (middle row), and 100 (bottom row). } 
	\label{Sierpinski_Vel}
\end{figure}

We first consider the rarefied gas flow through a two-dimensional cross section described by the Sierpinski carpet, which can be generated through recursion, beginning with a square. The square is cut into 9 congruent subsquares in a $3\times3$ grid, and the central subsquare is removed. The same procedure is then applied recursively to the remaining 8 subsquares. Resulting geometries after several levels of recursion are presented in Fig.~\ref{Sierpinski}.

Due to the symmetry, the one quarter of the level 1, 2, and 3 Sierpinski carpets  are divided into $60\times60$, $90\times90$, and $135\times135$ uniform cells, respectively. The molecular velocity space is represented by $32\times32\times12$ discrete grids. The iteration number, using $N=1.5$ in Eq.~\eqref{BGK_penlty2} for the hard-sphere gas, is always less than 55 for each rarefaction parameter, when the relative error in the mass flow rate $\mathcal{M}_P$ between two consecutive iteration steps is less than $10^{-5}$.

Figure~\ref{Sierpinski_Vel} displays the contour of flow velocities at different geometries and rarefaction parameters, where the characteristic flow length $\ell$ is chosen to be the side length of the largest square. When there is no solid inside the largest square (first column in Fig.~\ref{Sierpinski_Vel}), the maximum velocity is at the center of the domain. When there are some solids inside the largest square, when $\delta$ is small, it seems that large flow velocities are equally distributed near the central regions. However, when $\delta$ is large, large flow velocities are localized between the smallest squares which are far away from other larger squares, instead of in the central region of the carpets. This may be seen clearly in the Sierpinski carpet of level 3 (last column of Fig.~\ref{Sierpinski_Vel}). 

\begin{figure}[t]
	\centering
	{	\includegraphics[scale=0.5,viewport=5 0 460 340,clip=true]{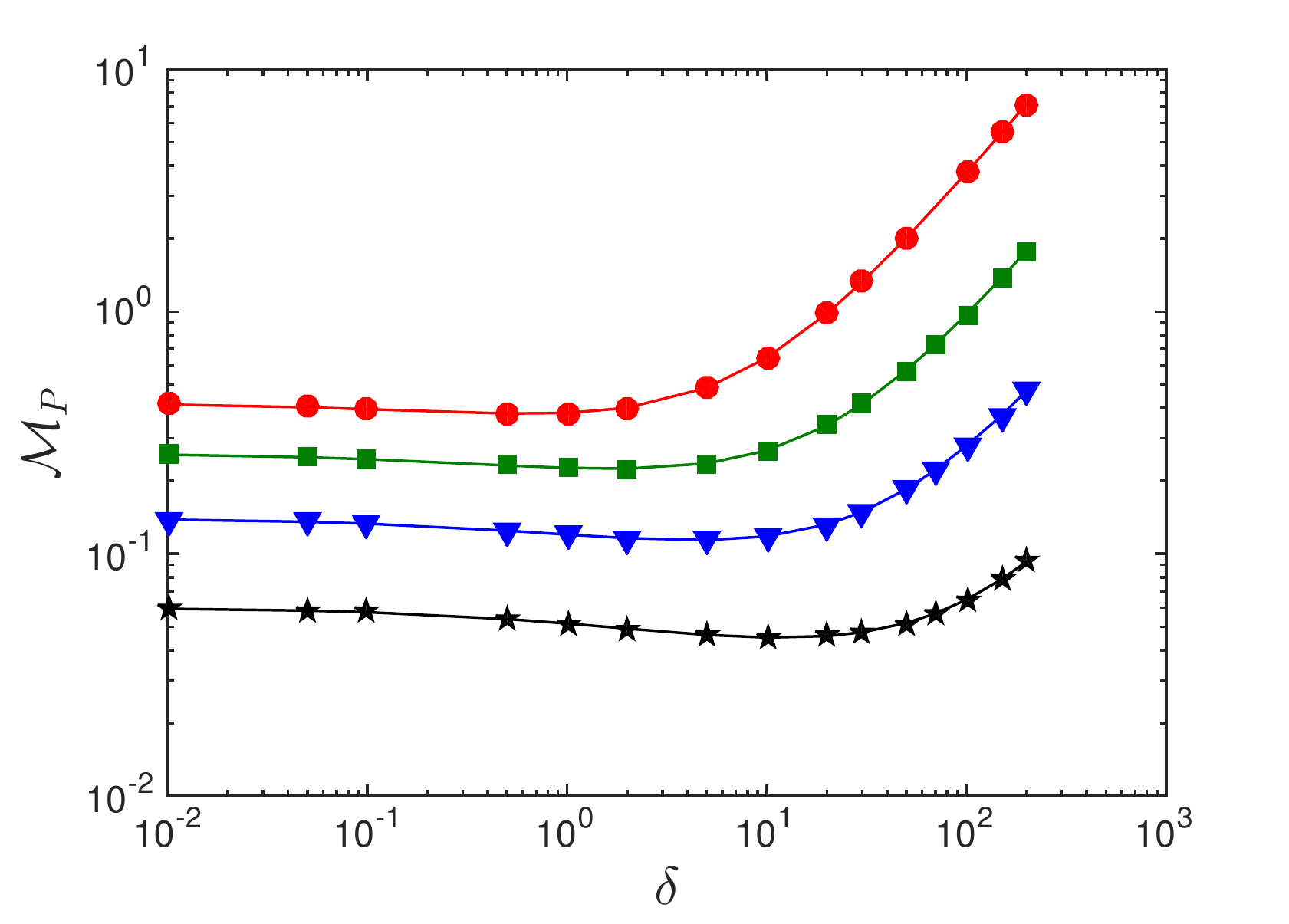}}
	{	\includegraphics[scale=0.5,viewport=5 0  460 340,clip=true]{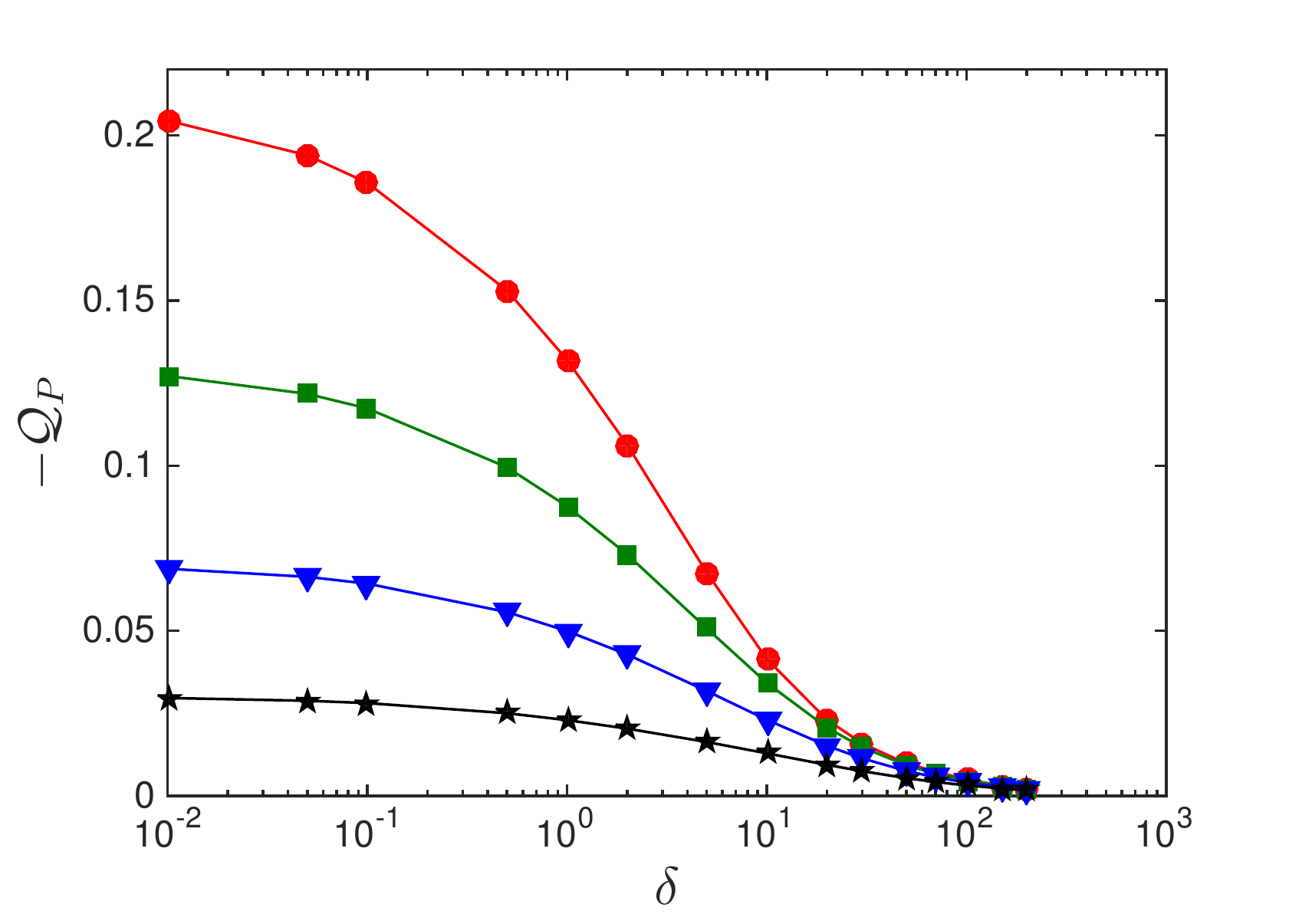}}
	\\
	\vskip 0.4cm
	{\includegraphics[scale=0.5,viewport=0 0 460 340,clip=true]{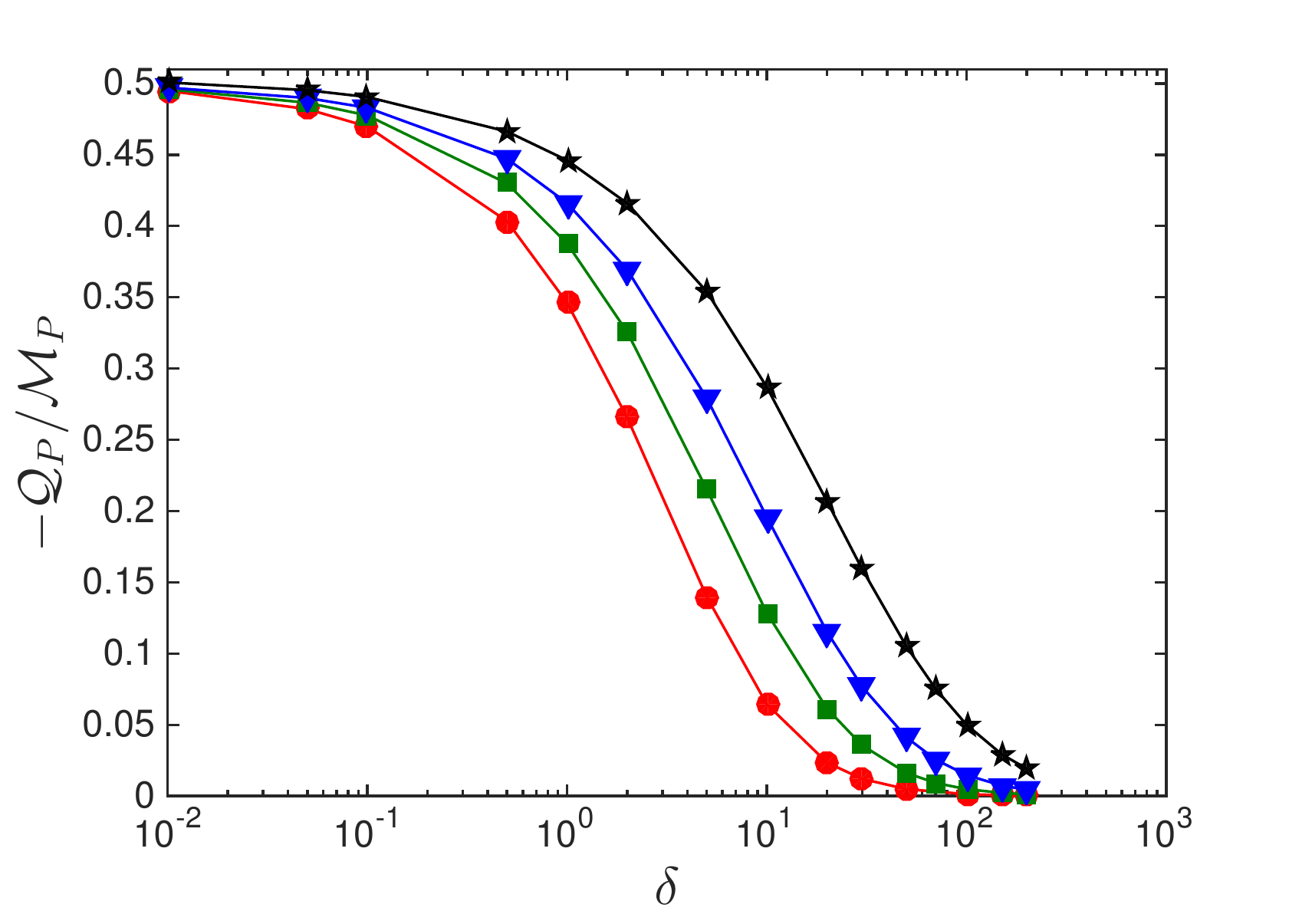}}
	\caption{Mass flow rate $\mathcal{M}_P$, heat flux flow rate $\mathcal{Q}_P$, and the thermomolecular pressure difference exponent $-\mathcal{Q}_P/\mathcal{M}_P$ in the Poiseuille flow of the hard-sphere gas through the Sierpinski carpets. Circles, squares, triangles, and pentagrams represent the results of the Sierpinski carpets generated at the recursion level 0, 1, 2, and 3, respectively.   } 
	\label{Sierpinski_mass}
\end{figure}

Figure~\ref{Sierpinski_mass} shows the flow rates in the Poiseuille flow of the hard-sphere gas through the Sierpinski carpet. The Knudsen minimum in the mass flow rate can be seen, however, the location of the minimum $\mathcal{M}_P$ shifts towards large values of $\delta$, as the recursion level of the Sierpinski carpet increases. This is because, in the calculation of $\delta$ according to Eq.~\eqref{delta}, the characteristic flow length $\ell$ is chosen to be the side length of the largest square, which is larger than, say, the smallest side length of the solids near which the flow velocity is maximum. As the recursion level increases, the porosity (the fraction of the void area) of the Sierpinski carpet decreases, so both the mass and heat flux flow rates decrease. We also plot in Fig.~\ref{Sierpinski_mass} the thermomolecular pressure difference exponent, which is an important parameter determining the performance of Knudsen pump. The exponent always increases with decreasing $\delta$ and approaches the value of 0.5 when $\delta\rightarrow0$ when the diffuse gas-surface boundary condition is used~\cite{SharipovCL3}. This may also indicate the correctness of our numerical simulations.

\subsection{Rarefied gas flow through random structures}

\begin{figure}[tp]
	\centering
	\subfloat[]{	\includegraphics[scale=0.7,viewport=50 30 260 240,clip=true]{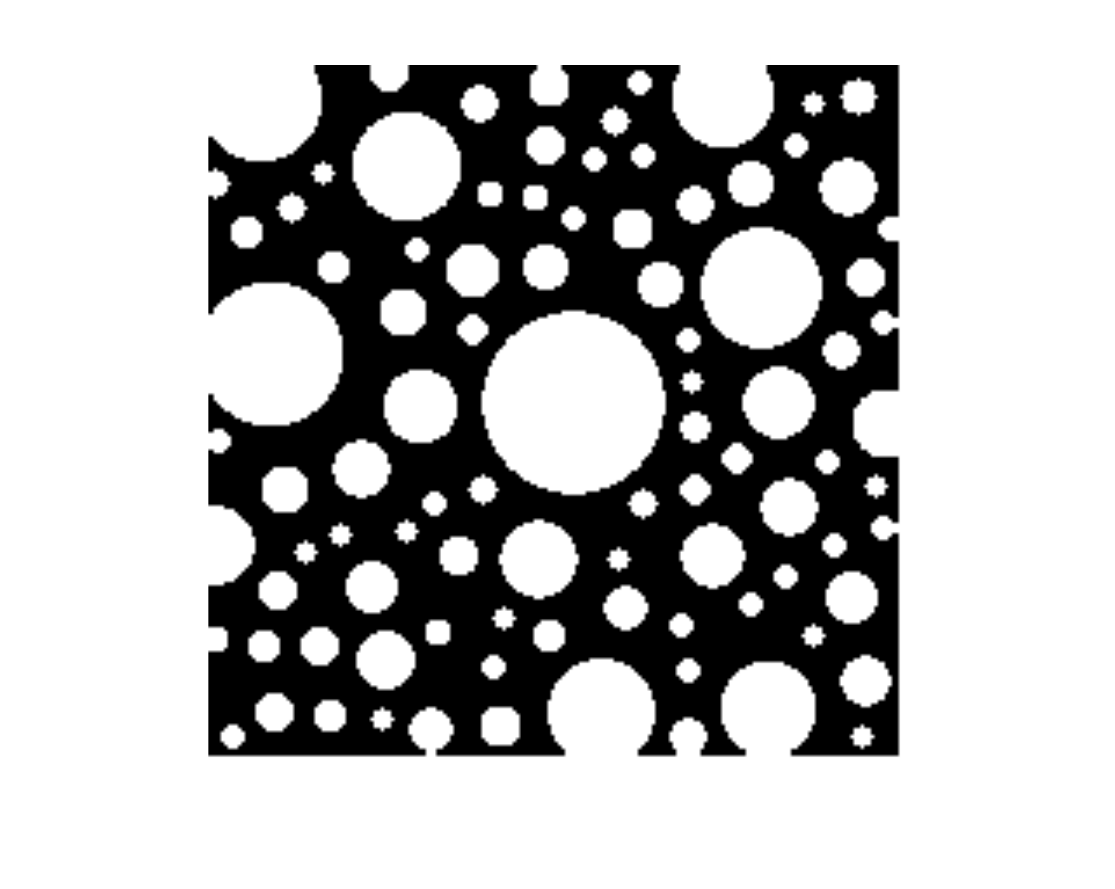}}
	\hskip 1.5cm
	\subfloat[]{	\includegraphics[scale=0.7,viewport=50 30 260 240,clip=true]{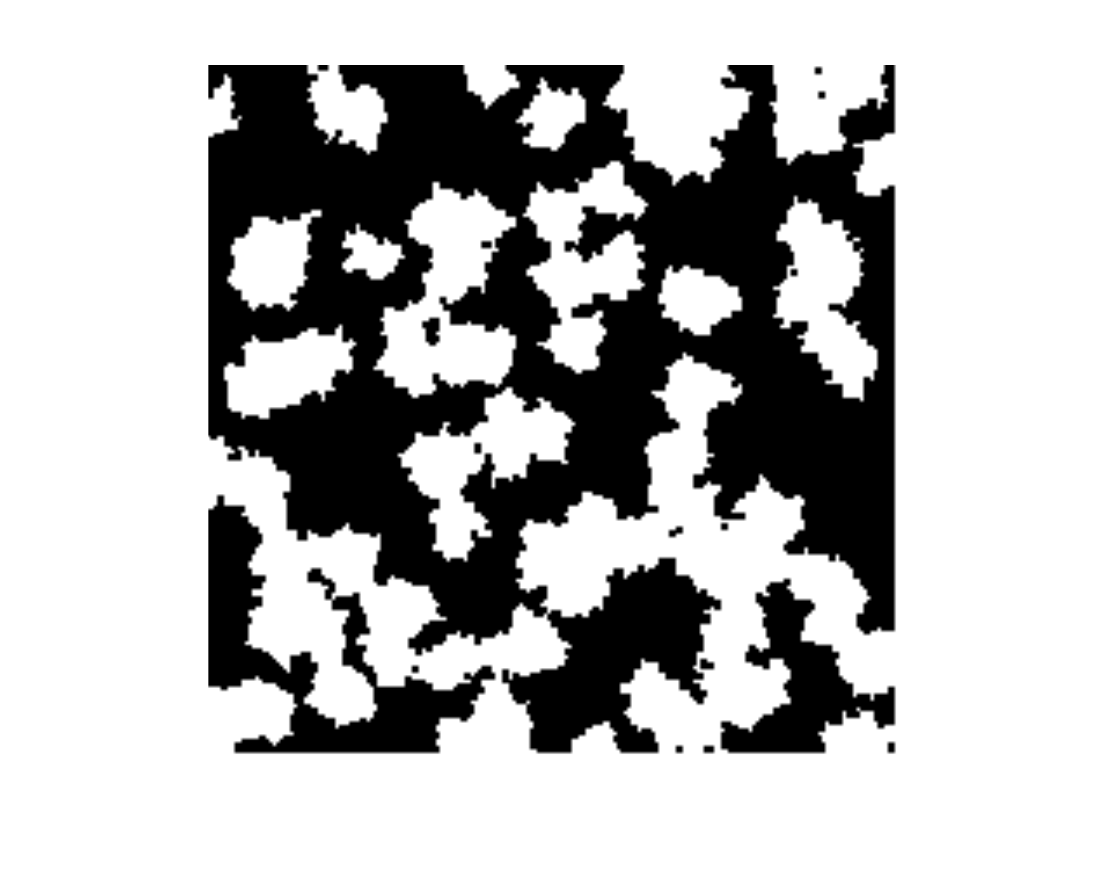}}
	\caption{Porous media with the porosity of 0.6, consisting of (a) discs of random position and radius, and (b) islands of different size and shape. White regions represent the solid, while the gas can flow through the black regions. } 
	\label{haihu}
\end{figure}

\begin{figure}[tp]
	\centering
	{	\includegraphics[scale=0.8,viewport=80 60 530 390,clip=true]{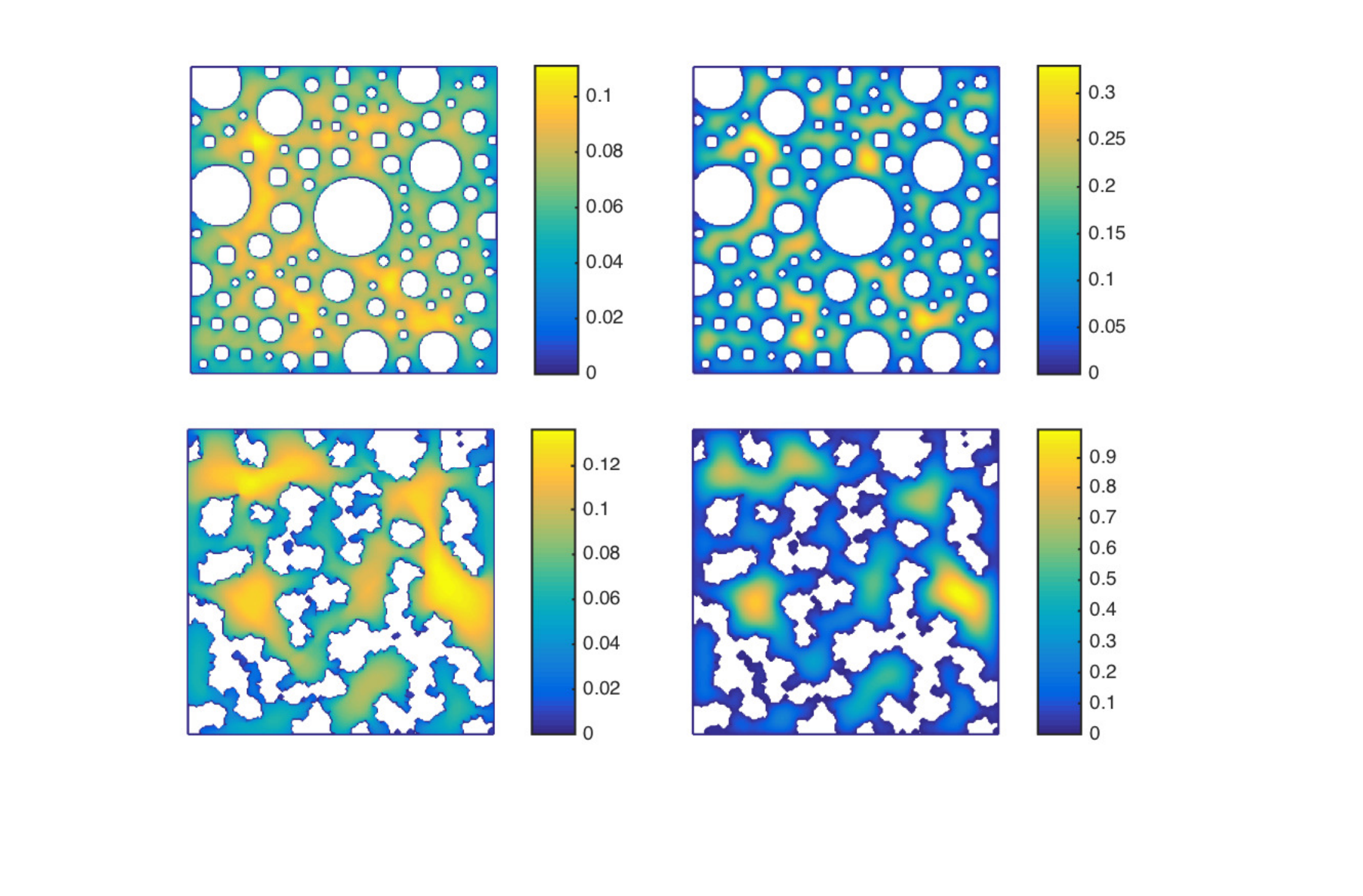}}
	\caption{Velocity contours in the Poiseuille flow of the hard-sphere gas through the two random porous media, when $\delta=0.01$ (left column) and 300 (right column). } 
	\label{Random_Vel}
\end{figure}

We then consider the rarefied gas flow through  two-dimensional porous media, where the porosity  is 0.6. The first porous medium is generated by adding circular solids of different radius randomly to a square. The radius ratio of the largest disc to the smallest is 10. The square is divided into $200\times200$ uniform cells, and the discs are approximated by the staircase, as visualized in Fig.~\ref{haihu}(a). The second porous medium, shown in Fig.~\ref{haihu}(b), also consisting of $200\times200$ uniform cells, is generated by the  quartet structure generation set~\cite{QSGS_Wang2007}.

\begin{figure}[t]
	\centering
	{	\includegraphics[scale=0.49,viewport=0 0 460 340,clip=true]{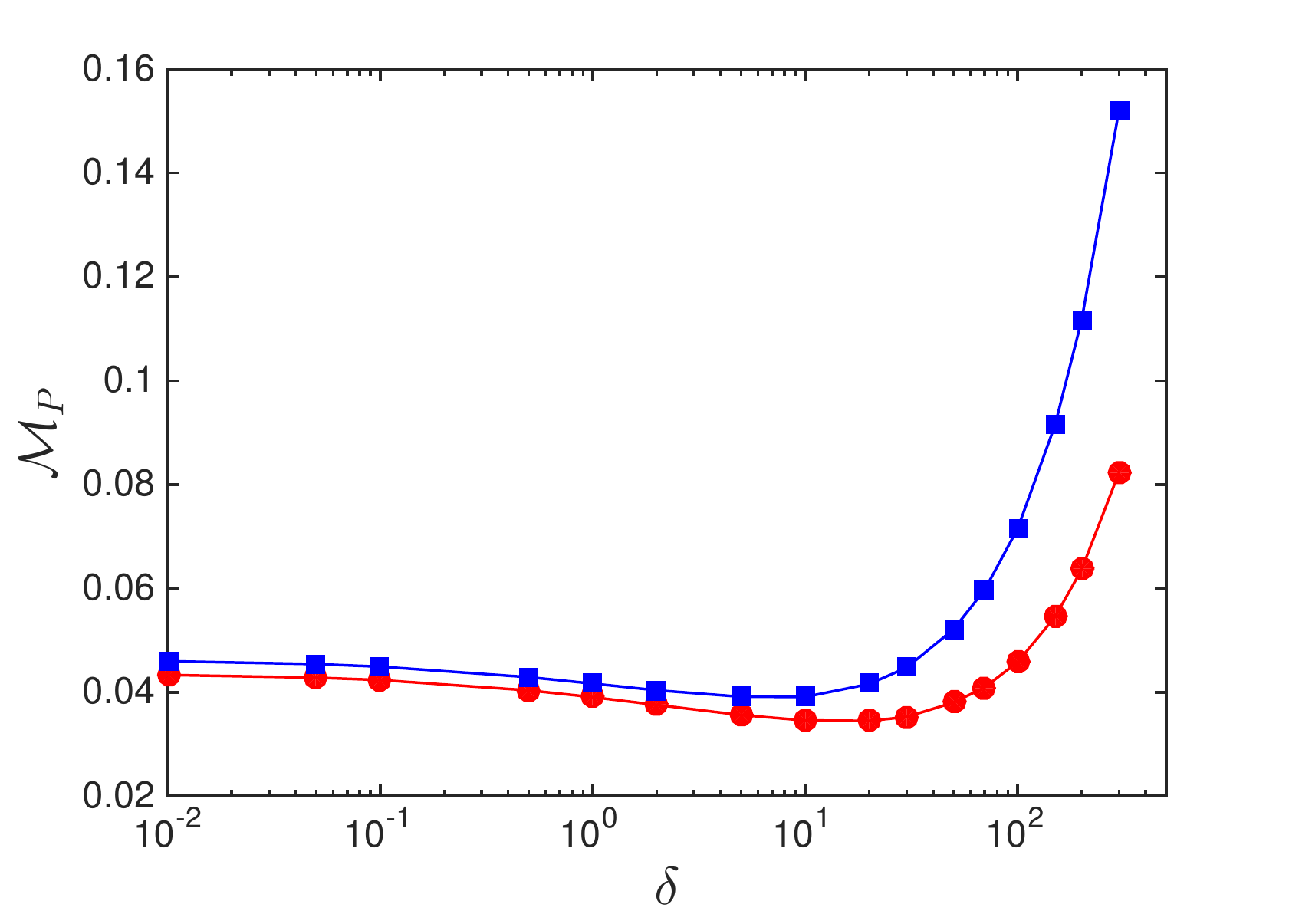}}
	{	\includegraphics[scale=0.49,viewport=0 0  460 340,clip=true]{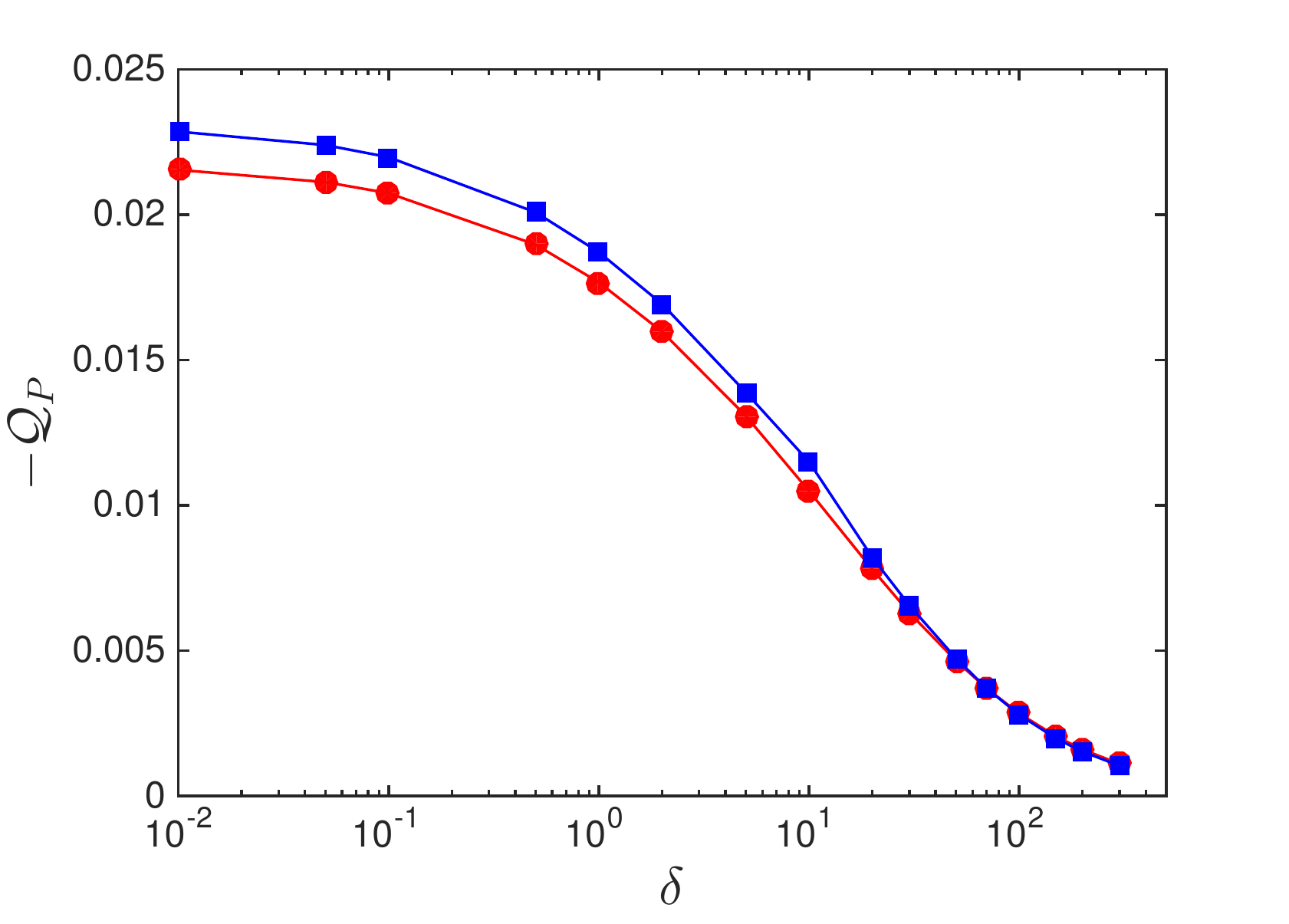}}
	\\
	\vskip 0.4cm
	{\includegraphics[scale=0.5,viewport=0 0 460 340,clip=true]{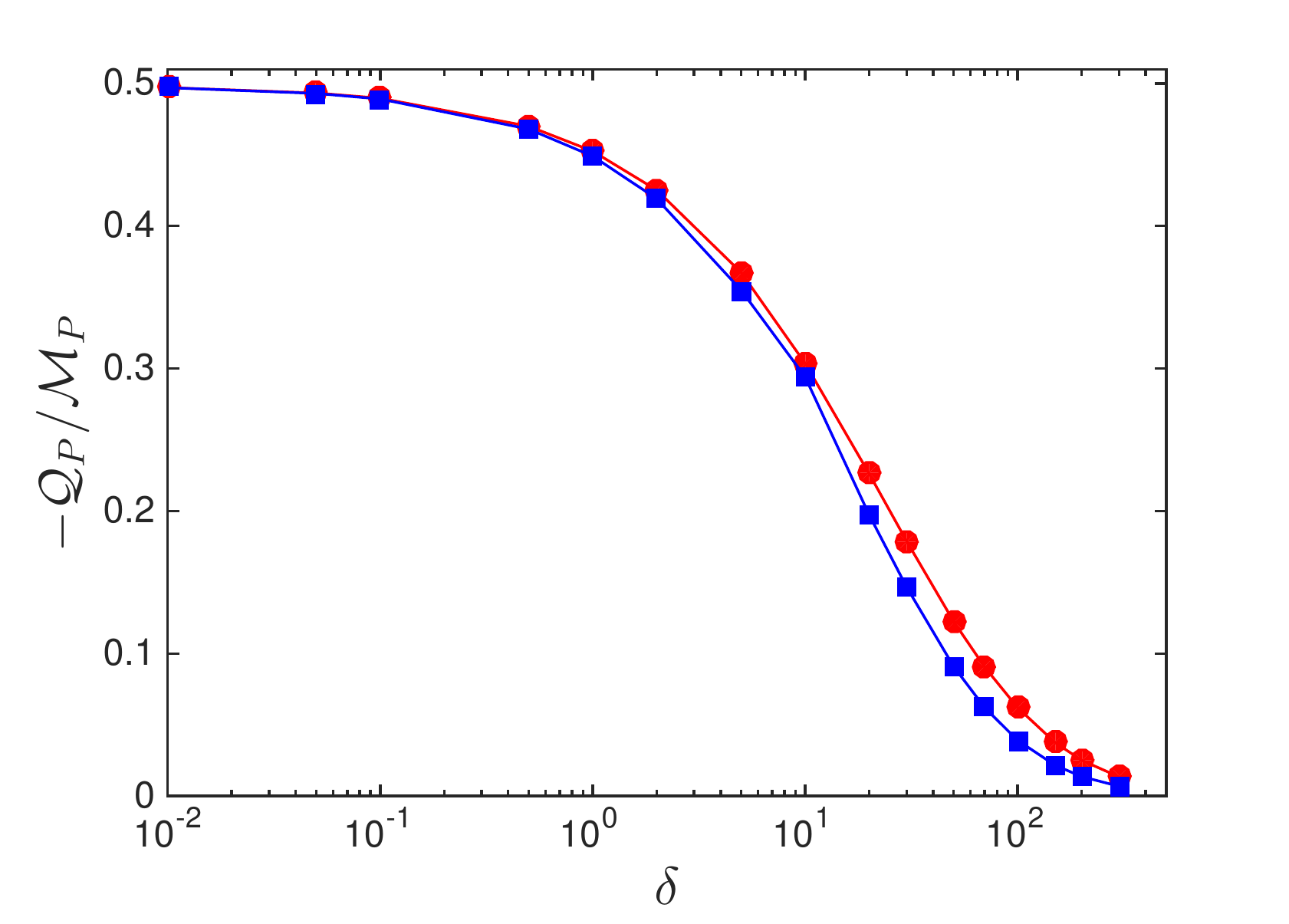}}
	\caption{Mass flow rate $\mathcal{M}_P$, heat flux flow rate $\mathcal{Q}_P$, and the thermomolecular pressure difference exponent $-\mathcal{Q}_P/\mathcal{M}_P$ in the Poiseuille flow of the hard-sphere gas through the Porous media consisting of random discs (circles) and islands (squares).   } 
	\label{Random_mass}
\end{figure}

Velocity contours are displayed in Fig.~\ref{Random_Vel} in the free molecular and near-continuum flow regimes, while the flow rates are shown in Fig.~\ref{Random_mass}. It is interesting to note that, in the free molecular flow regime, the mass flow rates of the two random porous media are nearly the same. However, in the near-continuum flow regime, the mass flow rate of the porous medium consisting of random squares is about twice larger than that of discs.

\section{A special case: SIS in the polar coordinate}\label{tuneSec}

The SIS developed above for the LBE works well in the Cartesian coordinates, for rarefied gas flows through general cross sections. For a flow through a circular cross section, the use of polar coordinates will save the computational cost significantly.

We consider the Poiseuille flow through a tube as an example, where the axis of the tube is along the $x_3$ direction, and its cross section is located in the $x_1$-$x_2$ plane. The spatial coordinates are normalized by the radius of the tube. Introduce the transformation $x_1=r\cos\theta$, $x_2=r\sin\theta$, $v_1=v_r\cos\theta$,  $v_2=v_r\sin\theta$, and define the VDF $h=h(r,\theta,v_r,v_3)$ in the cylindrical coordinates $v_r\in[0,+\infty)$, $\theta\in[0,2\pi]$, and $v_z\in(-\infty,\infty)$, the LBE can be written as:
\begin{equation}\label{polar}
v_1\frac{\partial h}{\partial r}-\frac{v_2}{r}\frac{\partial h}{\partial \theta}=L(h,f_{eq})+v_3f_{eq}. 
\end{equation}

To construct the SIS in the polar coordinates, an acceleration equation for the flow velocity $U_3(r)$ should be derived. Since the Laplace operator $\partial^2U_3/\partial x_1^2+\partial^2U_3/\partial x_2^2$ in Eq.~\eqref{diffusion_LBE} can be rewritten as $\frac{1}{r}\frac{\partial}{\partial {r}}\left(r\frac{\partial{U_3}}{\partial r}\right)$, our goal is to construct the acceleration equation in the following form:
\begin{equation}\label{polar_high}
\frac{1}{r}\frac{\partial}{\partial {r}}\left(r\frac{\partial{U_3}}{\partial r}\right)=-N\delta+\text{high-order terms},
\end{equation}
by taking the velocity moment of the LBE~\eqref{polar}.

Multiplying Eq.~\eqref{polar} by $2v_3$ and integrating over the molecular velocity space, we obtain the equation for the shear stress $P_{rz}=\int 2v_3{v_1h}d\textbf{v}=\int 2v_3{v_r^2\cos\theta}hdv_r{dv_3}d\theta$ as 
\begin{equation}\label{polar_stress}
\frac{1}{r}\frac{\partial{}}{\partial {r}}(rP_{rz}) =1.
\end{equation}

Multiplying Eq.~\eqref{polar} by $v_3v_1$, penalizing the linearized Boltzmann collision operator in the form of Eq.~\eqref{BGK_penlty2}, and integrating over the molecular velocity space, we obtain 
\begin{equation}
\frac{\partial}{\partial {r}}\int  v_3{v_1^2}hd\textbf{v}+\frac{1}{r}\int  v_3(v_1^2-v_2^2)hd\textbf{v}=-\frac{N\delta{}P_{rz}}{2}+\int{v_3v_1(L-NL_{BGK})}d\textbf{v},
\end{equation}
which is simplified, with the help of Eq.~\eqref{polar_stress}, into
\begin{equation}\label{polar_med}
\begin{split}
\frac{1}{r}\frac{\partial}{\partial {r}}  \bigg[r\frac{\partial}{\partial {r}}\int  2v_3{v_1^2}hd\textbf{v}+\int  2v_3(v_1^2-v_2^2)hd\textbf{v}-r\int{2v_3v_1(L-NL_{BGK})}d\textbf{v}\bigg]=-N\delta.
\end{split}
\end{equation}

If we express 
$\int{}2v_3{v_1^2}hd\textbf{v}=\int  v_3(2v_1^2-1)hd\textbf{v}+U_3$, the acceleration equation in the form of Eq.~\eqref{polar_high} can be derived. But for the practical numerical calculation,  the following first-order ordinary differential equation for the flow velocity is desirable:
\begin{equation}\label{polar_final}
\begin{split}
\frac{\partial{U_3}}{\partial r}=-\frac{N\delta{r}}{2}-\frac{\partial}{\partial {r}} \int  v_3{(2v_1^2-1)}hd\textbf{v}-\frac{1}{r}\int  2v_3(v_1^2-v_2^2)hd\textbf{v} +\int{2v_3v_1(L-NL_{BGK})}d\textbf{v}.
\end{split}
\end{equation}

In the numerical simulation, the spatial coordinate $r\in(0,1]$ is discretized by 150 nonuniform points, with most of the points located near the surface. Due to the symmetry, the truncated velocity $v_r\in(0,4)$ is discretized by 22 nonuniform points, with most of the points located near $v_r=0$, while $\theta\in[0,\pi]$ and $v_z\in(0,6)$ are discretized by 40 and 12 uniform points, respectively. The linearized Boltzmann collision operator is approximated by the fast spectral method: first, the spectrum of the VDF is calculated by Fourier transform from the cylindrical molecular velocity space to the Cartesian frequency space. Second, the fast spectral method~\cite{lei_Jfm} is applied to find the spectrum of the linearized Boltzmann collision operator in the Cartesian coordinate. Finally, the inverse Fourier transform is used to find the collision operator in the cylindrical space. As usual, the spatial derivatives in the LBE~\eqref{polar} are approximated by the second-order upwind finite difference, and Eq.~\eqref{polar_final} is solved by the implicit iterative scheme, while the acceleration equation is used to expedite the convergence to the steady-state solution, which is also solved using the second-order upwind finite difference, from $r=1$ to 0. Having obtained $U_3(r)$ from Eq.~\eqref{polar_final}, a correction in the VDF is performed, see Eq.~\eqref{guided}.

The numerical results for the mass and heat flux flow rates of the Poiseuille flow of the hard-sphere gas through a tube are summarized in Table~\ref{table_poiseuille_tube_compare}. It can be seen, as the rarefied gas flow through two parallel plates and rectangular cross sections, the SIS in the polar coordinates is also very efficient.

\begin{table}
	\centering
	\caption{Mass/heat flow rates in Poiseuille flow of hard-sphere and Maxwell molecules along a tube, as well as the number of iterations (Itr) to reach the convergence criterion $\epsilon=10^{-10}$ in the SIS. We choose $N=1$ in Eq.~\eqref{polar_final}. }
	\begin{tabular}{clcclcccccc}
		\hline
		&  \multicolumn{3}{l}{Hard-sphere molecules} 
		&  \multicolumn{3}{l}{Maxwell molecules}   \\
		\cline{2-7}
		$\delta$ & Itr & $\mathcal{M}_P$ & $-\mathcal{Q}_P$ & Itr & $\mathcal{M}_P$  & $-\mathcal{Q}_P$  \\

		$0.0$ & 3  & 0.752 & 0.376  & 3  & 0.752 & 0.376    \\
		
		$0.01$ & 6  & 0.736 & 0.362 & 7   & 0.731 & 0.355   \\
	
		$0.1$  & 11 & 0.699 & 0.318 & 15  & 0.693 & 0.307   \\

		$0.5$  & 24 & 0.690 & 0.247 & 34  & 0.692 & 0.244   \\
	
		$1$    & 33 & 0.723 & 0.202 & 48  & 0.732 & 0.205   \\

		$5$    & 46 & 1.150 & 0.083 & 62  & 1.179 & 0.091   \\

		$10$   & 45 & 1.747 & 0.047 & 60  & 1.786 & 0.052   \\
		
		$20$   & 49 & 2.966 & 0.025 & 65  & 3.024 & 0.028   \\

		$30$   & 47 & 4.192 & 0.017 & 63  & 4.269 & 0.019   \\				
		
		$50$   & 45 & 6.649 & 0.010 & 60  & 6.765 & 0.012   \\
	
		$100$  & 42 & 12.80 & 0.005 & 56  & 13.01 & 0.006   \\
		\hline
	\end{tabular}\par \label{table_poiseuille_tube_compare} 
\end{table}

\section{SIS for the LBE of gas mixtures}\label{sectionIII}

In this section we develop a SIS for the LBE for binary gas mixtures. For simplicity, only Poiseuille flow is considered, but the method can be generalized to flows driven by temperature and concentration gradients.

Let $f_A$ and $f_B$ be, respectively, the VDFs of gas components A and B with molecular mass $m_A$ and $m_B$, and molar fraction $\chi_A$ and $\chi_B=1-\chi_A$. Introducing the equilibrium VDF (in which the velocity is normalized by the most probable speed $v_{mA}=\sqrt{2k_BT_0/m_A}$ of component A~\cite{Wu:2015fk}):
\begin{equation}\label{Maxwell}
f_{\alpha,eq}(\textbf{v})=\chi_{{\alpha}}\left(\frac{m_{\alpha}}{\pi{m_A} }\right)^{3/2}\exp\left(-\frac{m_{\alpha}|\textbf{v}|^2}{2m_A}\right), \ \ \alpha=\text{A or B},
\end{equation}
and expressing the VDF in the form $f_{\alpha}=f_{\alpha,eq}+h_\alpha$, where $h_\alpha$ are deviated VDFs satisfying $|h_\alpha/f_{\alpha,eq}|\ll1$, the LBE for $h_\alpha$ reads
\begin{equation}\label{Boltzmann_lin}
\begin{split}
{v_1}\frac{\partial
	h_\alpha}{\partial x_1}+{v_2}\frac{\partial
	h_\alpha}{\partial x_2}=L_\alpha+S_\alpha,
\end{split}
\end{equation}
with the linearized Boltzmann collision operators  $L_\alpha=\sum_{\beta=1,2}Q_{\alpha\beta}(f_{\alpha,eq},h_\beta)+Q_{\alpha\beta}(h_\alpha,f_{\beta,eq})$, where the details of $Q_{\alpha\beta}$ can be found in Ref.~\cite{Wu:2015fk}. The source term for Poiseuille flow is
\begin{equation}
S_\alpha=-{X_P}v_3f_{\alpha,eq},
\end{equation}
and again we take $X_P=-1$.

When the deviated VDFs are known, the flow velocity normalized by $v_{mA}$ is calculated as $U_{\alpha}=\int h_{\alpha}v_{3}d\textbf{v}/\chi_{{\alpha}}$, shear stresses normalized by the total gas pressure $p_0$ are $P_{\alpha{13}}=2m_\alpha\int h_{\alpha}v_1v_3d\textbf{v}/m_A$ and $P_{\alpha{23}}=2m_\alpha\int h_{\alpha}v_2v_3d\textbf{v}/m_A$, and the heat flux normalized by $p_0v_{mA}$ is  $q_\alpha=\int{\left(m_\alpha|\textbf{v}|^2/m_A-5/2\right)}v_3hd\textbf{v}$. The dimensionless molecular flow rate $\mathcal{M}$ and heat flow rate $\mathcal{Q}$, normalized by the most probable speed of the gas mixture, are calculated as 
\begin{equation}
\begin{split}
\mathcal{M}_\alpha=\frac{1}{A}\sqrt{\frac{m}{m_A}}\iint U_\alpha dx_1dx_2, \\
\mathcal{Q}_\alpha=\frac{1}{A}\sqrt{\frac{m}{m_A}}\iint q_\alpha dx_1dx_2, 
\end{split}
\end{equation}
where $m=\chi_Am_A+(1-\chi_A)m_B$ is the average molecular mass of the mixture.

\subsection{The synthetic scheme for gas mixture}

As emphasized in the previous section, the relation ${\partial P_{13}}/{\partial x_1}+{\partial P_{23}}/{\partial x_2}=1$ is important to developing the SIS. For binary mixtures, this relation still holds, but now shear stresses are replaced by mixture shear stresses, i.e. $P_{13}=P_{A13}+P_{B13}$ and $P_{23}=P_{A23}+P_{B23}$. This poses an additional difficulty.

Following the basic steps in developing the synthetic equation, we obtain the following two equations for the flow velocity of each component:
\begin{equation}\label{LBE2_false}
\begin{aligned}[b]
\chi_\alpha\left( \frac{\partial^2 U_\alpha}{\partial x_1^2}+\frac{\partial^2 U_\alpha}{\partial x_2^2} \right)=&2\frac{\partial }{\partial x_1}\int v_1v_3L_\alpha{}d\textbf{v}+2\frac{\partial }{\partial x_2}\int v_2v_3L_\alpha{}d\textbf{v} \\
&
-\frac{1}{4}\left(\frac{\partial^2 F^\alpha_{2,0,1}}{\partial x_1^2}+2\frac{\partial^2 F^\alpha_{1,1,1}}{\partial x_1\partial x_2}+\frac{\partial^2 F^\alpha_{0,2,1}}{\partial x_2^2}\right),
\end{aligned}
\end{equation}
where $F^\alpha_{m,n,l}=\int h_\alpha{}H_m(v_1)H_n(v_2)H_l(v_3) d\textbf{v}$.

To obtain diffusion equations which recover the Stokes equation (an important step, guaranteeing fast convergence) in the hydrodynamic regime, we rewrite the linearized collision operator as\footnote{This is because the linearized Boltzmann collision operator for single-species gases can also be penalized into the form of $L=(L+N\delta{h})-N\delta{h}$, instead of the form given in Eq.~\eqref{BGK_penlty2}. The virtue of Eq.~\eqref{LBE_penlty} is that the resulting diffusion equations will be very simple (compared to those in Refs.~\cite{LajosSzalmas2016,Szalmas2010V} for the McCormack kinetic model), but without any loss of efficiency of the SIS.}:
\begin{equation}\label{LBE_penlty}
L_\alpha=(L_\alpha+N_\alpha\delta{h_\alpha})-N_\alpha\delta{h_\alpha},
\end{equation} 
where $N_\alpha$ are two constants, and let $2\int v_1v_3L_\alpha{}d\textbf{v}=\int v_1v_3(L_\alpha+N_\alpha\delta{h_\alpha}){}d\textbf{v}-N_\alpha\delta{m_AP_{\alpha13}/m_\alpha}$ and $2\int v_2v_3L_\alpha{}d\textbf{v}=\int v_2v_3(L_\alpha+N_\alpha\delta{h_\alpha}){}d\textbf{v}-N_\alpha\delta{m_AP_{\alpha23}/m_\alpha}$. Then, for component B, Eq.~\eqref{LBE2_false} is transformed to 
\begin{equation}\label{LBE2_true2}
\begin{aligned}[b]
\chi_B\left( \frac{\partial^2 U_B}{\partial x_1^2}+\frac{\partial^2 U_B}{\partial x_2^2} \right)=&
-N_B\delta\frac{m_A}{m_B}\left(\frac{\partial{P_{B13}}}{{\partial x_1}}+\frac{\partial{P_{B23}}}{{\partial x_2}}\right)-\frac{1}{4}\left(\frac{\partial^2 F^B_{2,0,1}}{\partial x_1^2}+2\frac{\partial^2 F^B_{1,1,1}}{\partial x_1\partial x_2}+\frac{\partial^2 F^B_{0,2,1}}{\partial x_2^2}\right) \\
&+2\frac{\partial }{\partial x_1}\int v_1v_3(L_B-N_B\delta{f_B})d\textbf{v}+2\frac{\partial }{\partial x_2}\int v_2v_3(L_B-N_B\delta{f_B})d\textbf{v},
\end{aligned}
\end{equation}
while for the component A, by using the relation ${\partial P_{13}}/{\partial x_1}+{\partial P_{23}}/{\partial x_2}=1$ with $P_{13}=P_{A13}+P_{B13}$ and $P_{23}=P_{A23}+P_{B23}$,  Eq.~\eqref{LBE2_false} is transformed to 
\begin{equation}\label{LBE2_true1}
\begin{aligned}[b]
\chi_A\left( \frac{\partial^2 U_A}{\partial x_1^2}+\frac{\partial^2 U_A}{\partial x_2^2} \right)=&
N_A\delta\left(\frac{\partial{P_{B13}}}{{\partial x_1}}+\frac{\partial{P_{B23}}}{{\partial x_2}}-1\right)-\frac{1}{4}\left(\frac{\partial^2 F^A_{2,0,1}}{\partial x_1^2}+2\frac{\partial^2 F^A_{1,1,1}}{\partial x_1\partial x_2}+\frac{\partial^2 F^A_{0,2,1}}{\partial x_2^2}\right) \\
&+2\frac{\partial }{\partial x_1}\int v_1v_3(L_A-N_A\delta{f_A})d\textbf{v}+2\frac{\partial }{\partial x_2}\int v_2v_3(L_A-N_A\delta{f_A})d\textbf{v}.
\end{aligned}
\end{equation}

\begin{figure}[tbp]
	\centering
	\includegraphics[scale=0.55,viewport=10 0 580 380,clip=true]{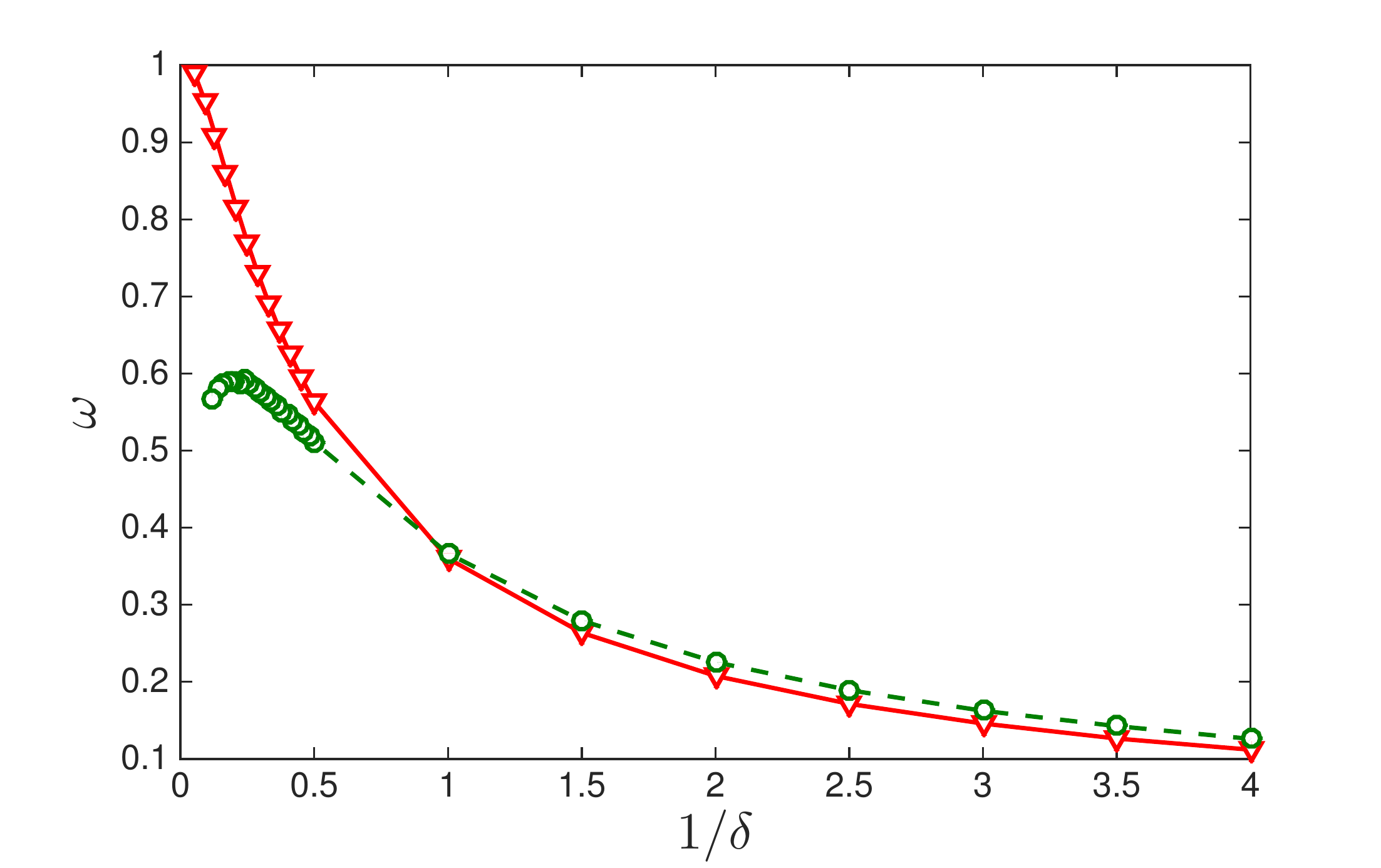}
	\caption{Eigenvalue $\omega$ versus the rarefaction parameter $\delta$ for the SIS (circles) and the CIS (triangles) of the LBE of an equimolar Ne-Ar mixture. } 
	\label{eig_mixture}
\end{figure}

We therefore propose the SIS for the LBE for binary gas mixtures: while the VDFs in Eq.~\eqref{Boltzmann_lin} are first solved by the CIS~\cite{Wu:2015fk}, flow velocities are updated according to diffusion equations~\eqref{LBE2_true2} and~\eqref{LBE2_true1}. Then the VDFs are corrected in a way similar to Eq.~\eqref{guided}. Note that the fastest convergence can be achieved when
\begin{equation}
N_\alpha=\frac{\int\nu_{\alpha,eq}(\textbf{v})f_{\alpha,eq}(\textbf{v})d\textbf{v}}{\delta\chi_\alpha},
\end{equation}
where $\nu_{\alpha,eq}$ is the equilibrium collision frequency of the $\alpha$-component.

The present SIS is readily generalized to multiple gas mixtures. Suppose there are $j$ gas components. For the velocity of the first component, the term $\partial{P_{B13}}/{\partial x_1}+\partial{P_{B23}}/\partial x_2$ in the diffusion equation~\eqref{LBE2_true1} can be replace by $\sum_{i=2}^j(\partial{P_{i13}}/{\partial x_1}+\partial{P_{i23}}/\partial x_2)$, while the diffusion equations for the flow velocities of the other components remain the same as Eq.~\eqref{LBE2_true2}, i.e. by replacing $B$ with the component index $i$. This method can also be applied to the McCormack kinetic equation~\cite{McCormack1973} for multiple gas mixtures, by simply replacing $L_\alpha$ in Eqs.~\eqref{LBE2_true2} and~\eqref{LBE2_true1} with that in the McCormack model; the resulting diffusion equations will be much simpler than those in Refs.~\cite{LajosSzalmas2016,Szalmas2010V}.

\subsection{Convergence analysis}

To show the efficiency of the proposed SIS for binary gas mixtures, we calculate the eigenvalue of the iteration as a function of the rarefaction parameter. The numerical procedure is essentially the same as that in Sec.~\ref{conv_analysis} for single-species gases. Fig.~\ref{eig_mixture} shows the eigenvalue of both the SIS and  CIS for an equimolar Ne-Ar mixture, where Ne and Ar are treated as hard-sphere molecules with a molecular diameter ratio of 0.711. It is clear that the SIS speeds up the slow convergence at large values of the rarefaction parameter. Also, when compared to the SIS for a single-species hard-sphere gas, the synthetic scheme for a binary gas mixture has roughly the same maximum eigenvalue, i.e. $\omega\simeq0.6$. So it is expected that the synthetic scheme for a binary gas mixture will be as efficient as that for a single-species gas. It is also interesting to noted that at small values of $\delta$ the eigenvalue of the SIS is slightly higher than that of the CIS; the reason for this is not clear. However, it will not affect the efficiency of the SIS because at small values of $\delta$ both the SIS and CIS converge rapidly, i.e. within a small number of iterations.

\begin{sidewaystable}
	\centering
	\caption{Molecular and heat flow rates in the Poiseuille flow of the Ne-Ar mixture between two parallel plates, alongside the number of iterations (Itr) to reach the convergence criterion $\epsilon=10^{-10}$ in the SIS. The LBE with the hard-sphere model is used. }
	\begin{tabular}{clcccclcccclcccccc}
		\hline
		&  \multicolumn{5}{l}{$\chi_{Ne}=0.1$} 
		&  \multicolumn{5}{l}{$\chi_{Ne}=0.5$}  
		&  \multicolumn{5}{l}{$\chi_{Ne}=0.9$}   \\  \cline{2-16} \\
		$\delta$ 
		& Itr & $\mathcal{M}_{Ne}$ & $\mathcal{M}_{Ar}$ & $-\frac{\mathcal{Q}_{Ne}}{\chi_{Ne}}$ & $-\frac{\mathcal{Q}_{Ar}}{\chi_{Ar}}$ 
		& Itr & $\mathcal{M}_{Ne}$ & $\mathcal{M}_{Ar}$ & $-\frac{\mathcal{Q}_{Ne}}{\chi_{Ne}}$ & $-\frac{\mathcal{Q}_{Ar}}{\chi_{Ar}}$ 
		& Itr & $\mathcal{M}_{Ne}$ & $\mathcal{M}_{Ar}$  & $-\frac{\mathcal{Q}_{Ne}}{\chi_{Ne}}$ & $-\frac{\mathcal{Q}_{Ar}}{\chi_{Ar}}$ \\

		$0.01$& 9 & 2.131& 1.416& 0.999& 0.639& 9 & 1.858& 1.227& 0.861& 0.544& 9 & 1.547& 1.016& 0.705& 0.440\\
		
		$0.05$& 12& 1.593& 1.071& 0.720& 0.459& 13& 1.393& 0.936& 0.619& 0.391& 13& 1.167& 0.787& 0.506& 0.316\\
	
		$0.1$ & 15& 1.386& 0.951& 0.607& 0.391& 15& 1.218& 0.840& 0.522& 0.334& 16& 1.030& 0.717& 0.428& 0.272\\
		
		$0.5$ & 31& 1.005& 0.766& 0.372& 0.251& 32& 0.912& 0.709& 0.323& 0.218& 32& 0.808& 0.647& 0.269& 0.180\\
	
		$1$   & 38& 0.909& 0.744& 0.284& 0.196& 38& 0.846& 0.707& 0.248& 0.170& 35& 0.773& 0.666& 0.208& 0.141\\
		
		$2$   & 47& 0.879& 0.776& 0.204& 0.142& 47& 0.841& 0.756& 0.179& 0.123& 43& 0.794& 0.730& 0.151& 0.102\\
	
		$5$   & 52& 1.016& 0.968& 0.115& 0.080& 48& 1.002& 0.964& 0.101& 0.068& 54& 0.977& 0.949& 0.085& 0.056\\
		
		$10$  & 51& 1.371& 1.346& 0.067& 0.045& 47& 1.367& 1.348& 0.058& 0.039& 62& 1.352& 1.337& 0.049& 0.031\\
		
		$20$  & 51& 2.153& 2.141& 0.036& 0.024& 46& 2.155& 2.145& 0.031& 0.021& 84& 2.145& 2.137& 0.026& 0.017  \\
		
		$30$  & 50& 2.955& 2.946& 0.024& 0.017& 45& 2.958& 2.951& 0.021& 0.014& 89& 2.950& 2.943& 0.018& 0.012  \\				
	
		$50$  & 49& 4.570& 4.565& 0.015& 0.010& 45& 4.574& 4.569& 0.013& 0.009& 94& 4.568& 4.562& 0.011& 0.007  \\
		
		$100$ & 49& 8.624& 8.620& 0.008& 0.006& 44& 8.625& 8.622& 0.007& 0.005& 98& 8.623& 8.614& 0.007& 0.004  \\
		\hline
	\end{tabular}\par \label{table_poiseuilleHF_1d_mixture} 
\end{sidewaystable}

\subsection{Numerical simulations of Poiseuille flow}

\begin{figure}[t]
	\centering
	\includegraphics[scale=0.6,viewport=40 10 800 500,clip=true]{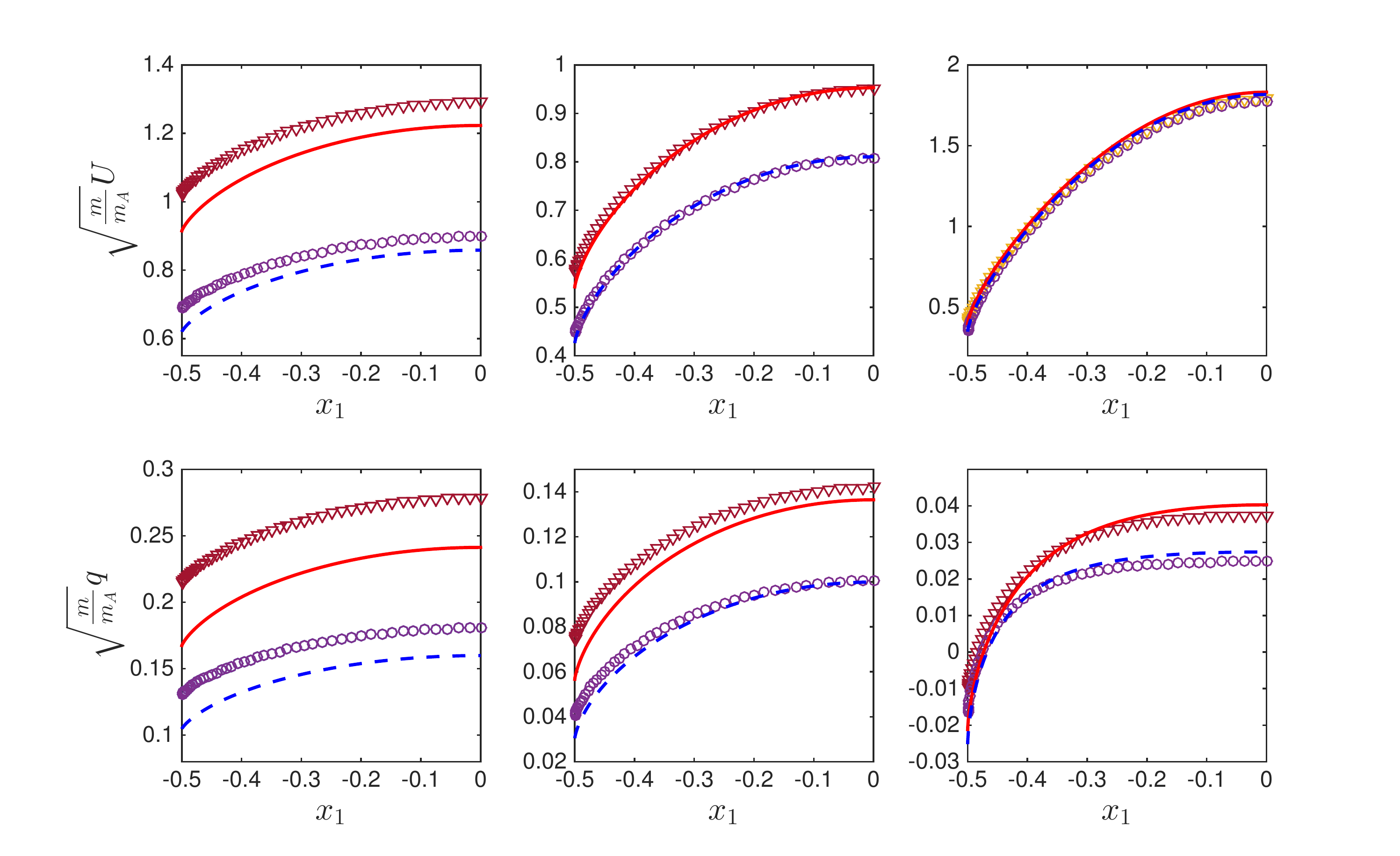}
	\caption{Velocity (top) and heat flux (bottom) profiles in the Poiseuille flow of an equimolar Ne-Ar mixture between two parallel plates, where $\delta=0.1$, 1, and 10 in the left, middle, and right column, respectively. Triangles and circles are the profiles of Ne and Ar, respectively, obtained from the LBE with the hard-sphere model, while solid and dashed lines are the corresponding profiles using the Lennard-Jones potential. } 
	\label{mixture_vel}
\end{figure}

Following the convergence analysis, numerical simulations are now conducted on the Poiseuille flow of a Ne-Ar mixture between two parallel plates. First, the hard-sphere potential is considered. Table~\ref{table_poiseuilleHF_1d_mixture} shows the molecular and heat flow rates of each component over a wide range of gas rarefaction, when the molar fraction of Ne is 0.1, 0.5, and 0.9. From this table we see that the SIS is efficient, since convergence is reached within 100 iterations when $N_\alpha=1$; when using $N_\alpha=1.5$ in Eqs.~\eqref{LBE2_true2} and~\eqref{LBE2_true1}, converged solutions can be obtained within 50 iterations. Such an efficient method enables the study of the flow dynamics in the whole range of gas rarefaction.

In the free molecular and transition regimes, the molecular flow rate of the lighter species is always larger than that of the heavier species, however, in the near-continuum regime the molecular flow rates are the same for both gas components. When compared to the results in Table~\ref{table_poiseuille_1d_compare} we see that the difference in the molecular flow rates is within 1\% in the near-continuum regime. A comparison of velocity and heat flux profiles in different flow regimes is visualized in Fig.~\ref{mixture_vel}, which shows that the difference between the velocity profile of each component is reduced as the rarefaction parameter increases. As $\delta$ increases, the difference between the heat flux divided by the molar fraction of each component also decreases, but in the near-continuum limit the heat flux of Ne is roughly $\sqrt{2}$ times of that of Ar, while the difference in the molecular flow rate goes to zero.

\begin{figure}[t]
	\centering
	\includegraphics[scale=0.55,viewport=20 50 600 560,clip=true]{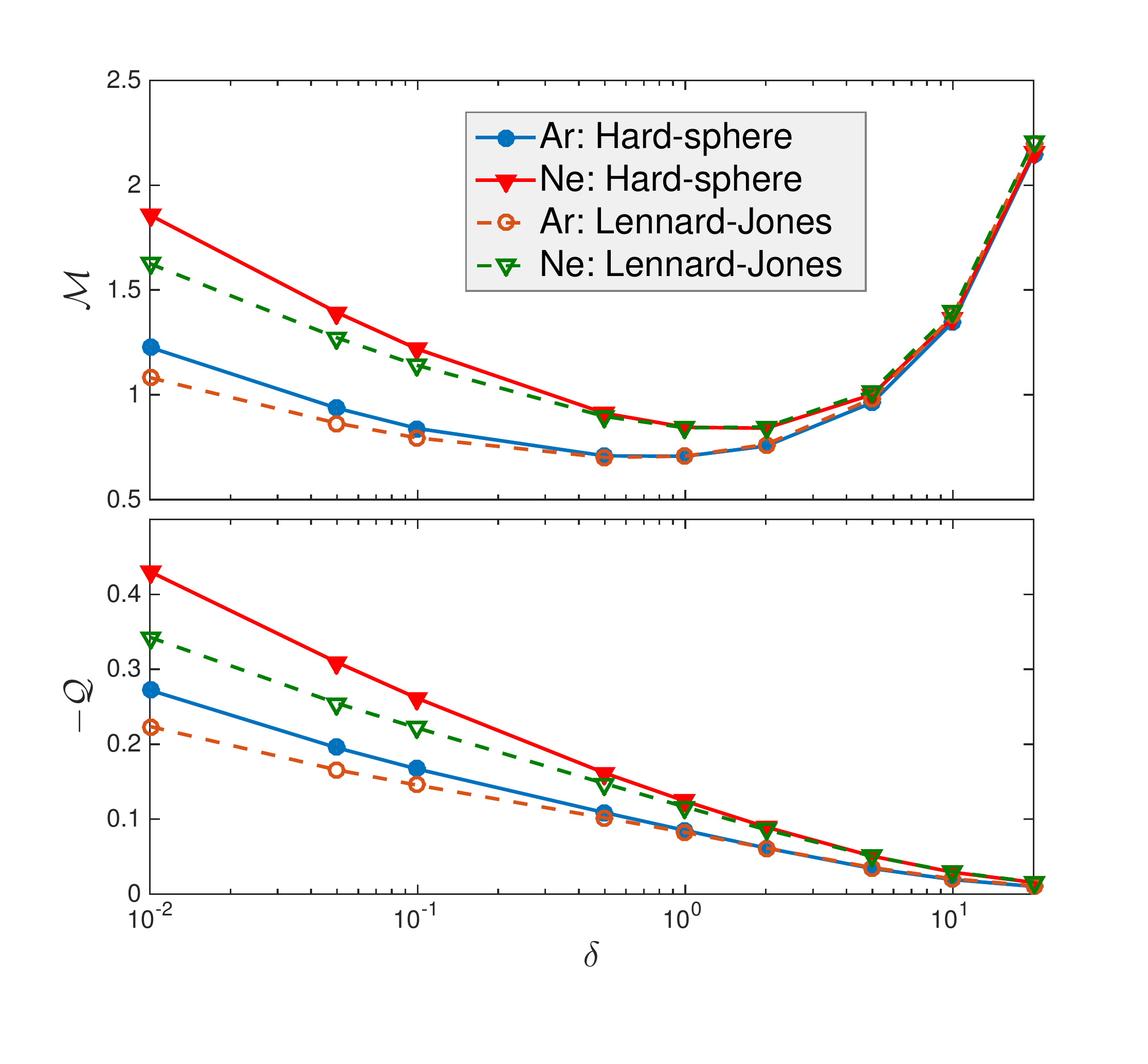}
	\caption{A comparison between the Lennard-Jones and hard-sphere potentials: molecular flow rate (top) and heat flow rate (bottom) profiles in the Poiseuille flow of an equimolar Ne-Ar mixture between two parallel plates. } 
	\label{mixture_mac}
\end{figure}

\begin{table}[tp]
	\centering
	\caption{Molecular and heat flow rates in Poiseuille flow of the equimolar Ne-Ar mixture through rectangular cross sections of aspect ratio one and two, as well as the iterative number (Itr) needed to reach the convergence criterion $\epsilon=10^{-7}$ in the SIS. $N_\alpha=1.5$ is adopted in Eqs.~\eqref{LBE2_true2} and~\eqref{LBE2_true1}. The LBE with the hard-sphere model is used.}
	\begin{tabular}{clcccclccccccccccc}
		\hline
		&  \multicolumn{5}{l}{Aspect ratio 1} 
		&  \multicolumn{5}{l}{Aspect ratio 2}   \\  \cline{2-11} 
		$\delta$ 
		& Itr & $\mathcal{M}_{Ne}$ & $\mathcal{M}_{Ar}$ & $-2{\mathcal{Q}_{Ne}}$ & $-2{\mathcal{Q}_{Ar}}$ & Itr & $\mathcal{M}_{Ne}$ & $\mathcal{M}_{Ar}$ & $-2{\mathcal{Q}_{Ne}}$ & $-2{\mathcal{Q}_{Ar}}$ \\

		$0.01$& 5 &  0.506  & 0.358  & 0.251  & 0.177  & 5  & 0.692 & 0.490 & 0.343 & 0.241\\
		
		$0.05$& 5 & 0.493   & 0.349  & 0.241  & 0.167  & 6  & 0.669 & 0.474 & 0.324 & 0.224\\
	
		$0.1$ & 6 & 0.483   & 0.344  & 0.232  & 0.160  & 7  & 0.653 & 0.466 & 0.309 & 0.212\\
		
		$0.5$ & 11& 0.452   & 0.338  & 0.195  & 0.134  & 13 & 0.604 & 0.459 & 0.249 & 0.170\\

		$1$   & 13& 0.441   & 0.347  & 0.170  & 0.117  & 16 & 0.589 & 0.475 & 0.210 & 0.144\\
		
		$2$   & 17& 0.442   & 0.373  & 0.139  & 0.096  & 19 & 0.597 & 0.518 & 0.164 & 0.112\\
	
		$5$   & 19& 0.497   & 0.459  & 0.091  & 0.061  & 20 & 0.699 & 0.659 & 0.101 & 0.067\\
		
		$10$  & 19& 0.630   & 0.610  & 0.057  & 0.038  & 21 & 0.926 & 0.904 & 0.060 & 0.040\\
		
		$20$  & 23& 0.928   & 0.917  & 0.032  & 0.021  & 23 & 1.417 & 1.406 & 0.033 & 0.022\\
		
		$30$  & 23& 1.235   & 1.228  & 0.022  & 0.015  & 23 & 1.920 & 1.913 & 0.023 & 0.015\\

		$50$  & 25& 1.856   & 1.852  & 0.014  & 0.009  & 25 & 2.932 & 2.929 & 0.014 & 0.009\\
		
		$100$ & 30& 3.416   & 3.416  & 0.007  & 0.005  & 30 & 5.474 & 5.474 & 0.007 & 0.005\\
		\hline
	\end{tabular}\par \label{table_poiseuilleHF_2d_mixture} 
\end{table}

We have also simulated the Ne-Ar mixture flow based on the LBE for Lennard-Jones potentials for the first time, where the fast spectral approximation of the Boltzmann collision operator is reported in Ref.~\cite{Wu:2015yu}. The influence of the molecular model on the velocity and heat flux profiles is shown in Fig.~\ref{mixture_vel}: the hard-sphere model overestimates the velocity and heat flux in the free-molecular regime, but in the transition and near-continuum regimes the difference between the two molecular models reduces as $\delta$ increases. This is in good agreement with the observations in the single-species case~\cite{Sharipov2009,Wu:2015yu}.

Figure~\ref{mixture_mac} shows the molecular and heat flow rates as a function of the rarefaction parameter. The two flow rates obtained from the LBE with Lennard-Jones potentials is smaller than that for the hard-sphere model when $\delta<1$ (i.e. when $\delta=0.01$, the particle flow rates are about 15\% smaller, while the heat flow rates are about 25\% smaller). This situation is reversed for the molecular flow rate, but differences between the two molecular models are nearly indistinguishable when $\delta>1$.

\begin{figure}[t]
	\centering
	\includegraphics[scale=0.69,viewport=50 40 590 560,clip=true]{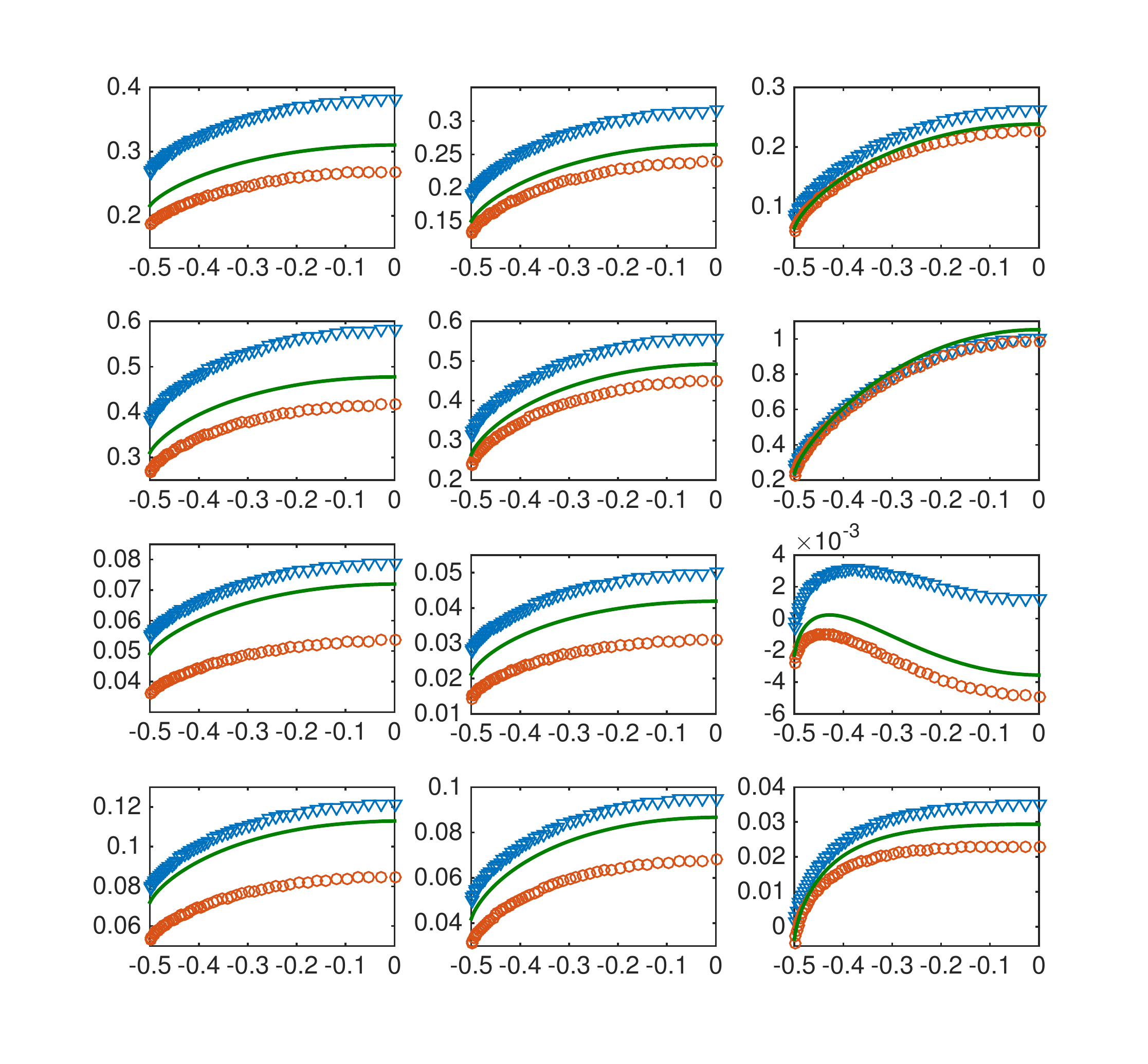}
	\caption{Velocity $(\sqrt{m/m_A}U)$ and heat flux $(\sqrt{m/m_A}q)$ profiles in the Poiseuille flow of an equimolar Ne-Ar mixture through a square cross section, where $\delta=0.1$, 1, and 10 in the left, middle, and right columns, respectively. The first and second (third and fourth) rows show the velocity (heat flux) along the boundary and the center line of the cross section, respectively. Triangles: Ne, circles: Ar, while solid lines in the first and second (third and fourth) rows show the velocity (half heat flux) obtained from the single-species LBE. } 
	\label{mixture2D}
\end{figure}

We now calculate the Poiseuille flow of an equimolar Ne-Ar mixture along channels of rectangular cross sections, based on the hard-sphere model. To the best of our knowledge, the LBE for a gas mixture has never been solved in a two-dimensional geometry before, because of the numerical complexity, but now the problem is tackled here by the SIS and the fast spectral method~\cite{Wu:2015fk}. The discretization of the spatial domain of a square cross section is the same as that in Sec.~\ref{NumSingle}, while for a rectangular cross section of aspect ratio 2 the spatial domain is discretized by $50\times100$ cells, and the characteristic length $\ell$ is chosen to be the shorter side. Table~\ref{table_poiseuilleHF_2d_mixture} summarizes the LBE solution for the molecular and heat flow rates in two-dimensional Poiseuille flow over a wide range of the rarefaction parameter for the first time. The iterations needed to achieve the convergence criterion $\epsilon=10^{-7}$ are fewer than 40 when $\delta\le100$, demonstrating the efficiency of the SIS. The normalized molecular flow rates of Ne and Ar through the rectangular cross section with aspect ratio 2 are always larger than those in the aspect ratio 1 case. However, although the heat flow rates through the rectangular cross section with an aspect ratio of 2 are larger than those with the aspect ratio 1 when $\delta<20$, they become roughly the same when $\delta\ge20$. Typical velocity and heat flux profiles of the Ne-Ar mixture through the square cross section in the free molecular, transition, and near-continuum regimes are shown in Fig.~\ref{mixture2D}, and are also compared to those of the single-species gas in the same geometry. From this figure we see that the velocity and heat flux of Ne is always larger than that of Ar, while the corresponding results for a single-species gas lie between these of Ne and Ar.

\section{Conclusions}\label{summary}

We have proposed a synthetic iterative scheme to accelerate the convergence of the linearized Boltzmann equation for gas flows driven by pressure and temperature gradients in long channels. By penalizing the linearized Boltzmann collision operator $L$ into the form of $L=(L+NL_{BGK})-NL_{BGK}$ or $L=(L+N\delta{h})-N\delta{h}$, a diffusion equation has been derived for the macroscopic flow velocity. The velocity distribution function in the linearized Boltzmann equation was first solved by the conventional iterative scheme, where the linearized Boltzmann collision operator was approximated by the fast spectral method. Then the flow velocity was obtained by solving the diffusion equation, which was finally used to correct the velocity distribution function. In this way the slow convergence of the conventional iterative scheme in the near-continuum flow regime has been tackled, as we found, through the numerical solution of Poiseuille and thermal transpiration flows, that the synthetic iterative scheme is faster than the conventional iterative scheme by up to several orders of magnitude.

The tuning parameter $N$ controls the convergence rate of the synthetic iterative scheme. While synthetic iterative schemes for other kinetic model equations (such as the linearized Bhatnagar-Gross-Krook model and the McCormack model) have been proposed with $N=1$, in numerical investigations we found that for the linearized Boltzmann equation the fastest convergence is achieved when $N$ roughly equals the ratio of the equilibrium collision frequency to the rarefaction parameter. Thus, $N$ varies with the intermolecular potential: for a single-species gas, we found that $N\approx1.5$ for the hard-sphere gas model and $N\approx2$ for the Maxwell gas model we used.

We also extended the synthetic iterative scheme to binary gas mixtures, and both the hard-sphere and Lennard-Jones potentials have been considered. As an example, Poiseuille flow of a Ne-Ar mixture was simulated in order to test the computational performance as well as the influence of the intermolecular potential. The synthetic iterative scheme required only a limited number of iterations over the whole range of gas rarefaction. Based on this efficient scheme, the Poiseuille flow of a Ne-Ar mixture between two parallel plates was simulated for the first time using the realistic Lennard-Jones potential. We found that the hard-sphere gas model overestimates the particle and heat flow rates when $\delta<1$. Poiseuille flow of a Ne-Ar mixture through two-dimensional rectangular cross sections was also simulated using the linearized Boltzmann equation for the first time. The molecular and heat flow rates were tabulated, and representative velocity and heat flux profiles were also shown for benchmarking.

The present method can be extended to the efficient calculation of flows of multiple gas mixtures. In particular, our method can be applied to the McCormack model and we believe that the resulting diffusion equations for the flow velocity of each component will be much simpler than those in Refs.~\cite{szalmas2010,szalmas2016}, while diffusion equations for the heat flux are not necessary since the linearized Boltzmann equation converges fast enough without accelerating the convergence of the heat flux. Our synthetic iterative method can also be applied straightforwardly to other canonical gas flows, such as Couette flow and the flow driven by a concentration gradient. However, it requires future work to investigate whether this method can be applied to other gas flow systems or not.

\section*{Acknowledgments}

This work is financially supported in the UK by the Engineering and Physical Sciences Research Council (EPSRC) under grants EP/M021475/1, EP/L00030X/1, EP/K038621/1, EP/I011927/1, and EP/N016602/1. H. Liu gratefully acknowledges the financial support of the ``Thousand Talents Program" for Distinguished Young Scholars and the National Natural Science Foundation of China under Grant No.~51506168. L. Wu acknowledges the financial support of an Early Career Researcher International Exchange Award from the Glasgow Research Partnership in Engineering.

\bibliographystyle{elsarticle-num}
\bibliography{Bib2}

\end{document}